\documentclass{eptcs}
\usepackage[utf8]{inputenc}
\usepackage{lhs}

\title{Foundations of Total Functional Data-Flow Programming}

\author{Baltasar Tranc{\'o}n y Widemann \institute{Ilmenau University of Technology, DE} \email{baltasar.trancon@tu-ilmenau.de} \and Markus
  Lepper \institute{\texttt{<semantics\kern1pt/>} GmbH, DE}}

\def\titlerunning{Foundations of Total Functional Data-Flow Programming}
\def\authorrunning{B.\ Tranc{\'o}n y Widemann \& M.\ Lepper}

\usepackage{amsmath,amssymb,stmaryrd}
\usepackage{semantic}
\usepackage[all,cmtip]{xy}
\usepackage{sigboxes}
\usepackage{microtype}
\usepackage{listings}

\usepackage{booktabs}
\usepackage{multirow}

\setnamespace{0pt}

\newcommand\bigsquig[1]{\leftarrow\kern-5pt\langle #1\rangle\kern-5pt\rightarrow}

\newcommand\insep{\mathbin/}
\newcommand\outsep{\mathbin/}

\newcommand\Person[1]{\textsc{#1}}

\newcommand\yes{\bullet}
\newcommand\some{\circ}
\newcommand\no{-}

\makeatletter
\newcommand\Power{\@ifstar{\mathbf{P}\kern-0.25ex_{+}}{\mathbf{P}}}
\newcommand\PowerI{\mathbf{P}\kern-0.25ex_{1}}
\renewcommand\Stream[1]{\mathbf{S}_{#1}}
\newcommand\Trans[1][\,]{\mathbf{T}_{#1}}
\newcommand\LTrans[1][\,]{\widehat{\mathbf{T}}_{#1}}

\newcommand\sem{\@ifstar{\sem@star}{\sem@nostar}}
\newcommand\sem@nostar[2][]{\sem@star{\mathbb{#1}}{#2}}
\newcommand\sem@star[2]{#1[\kern-0.33ex[#2]\kern-.33ex]}

\makeatother

\newcommand\ana[1]{[\kern-0.5ex(#1)\kern-0.5ex]}

\newcommand\causto{\stackrel{!}{\to}}

\def\Q#1#2.{\operatorname{\textstyle#1} #2 \mathpunct.}
\newcommand\rewrite[1][]{%
  ~\mathrel{\mathop{\Longmapsto}\limits^{\mathrm{#1}}}~}

\usepackage{xifthen}

\newcommand\quadri[4]{%
  \ifthenelse{\isempty{#1#2#3#4}}{}{%
    \ifthenelse{\isempty{#1#2}}{}{%
      \ifthenelse{\isempty{#1}}{}{%
        #1 \mathrel/
      } #2 \mathrel\rightarrow} #3
    \ifthenelse{\isempty{#4}}{}{\mathrel/ #4}}}

\newcommand\judge[6]{{} \quadri{#1}{#2}{#3}{#4}%
  \ifthenelse{\isempty{#5}}{}{{} \mathrel\Vert #5}{} \mathrel\vdash #6}

\newcommand\rjudge[5]{\judge{#1}{#2}{#3}{#4}{}{#5}}
\newcommand\ejudge[4]{\judge{#1}{}{#2}{#3}{}{#4}}
\newcommand\pjudge[3]{\judge{}{#1}{\begin{NEWENV4}#2\end{NEWENV4}}{}{}{#3}}

\newenvironment{NEWENV}{
  \ignorespaces}{\ignorespacesafterend}
\newenvironment{NEWENV2}{
  \ignorespaces}{\ignorespacesafterend}
\newenvironment{NEWENV3}{
  \ignorespaces}{\ignorespacesafterend}
\newenvironment{NEWENV4}{
  \ignorespaces}{\ignorespacesafterend}

\begin{document}
\maketitle

\begin{abstract}
  The field of declarative stream programming (discrete time, clocked
  synchronous, modular, data-centric) is divided between the data-flow
  graph paradigm favored by domain experts, and the functional
  reactive paradigm favored by academics.  In this paper, we describe
  the foundations of a framework for unifying functional and data-flow
  styles that differs from FRP proper in significant ways: It is based
  on set theory to match the expectations of domain experts, and the
  two paradigms are reduced symmetrically to a low-level middle
  ground, with strongly compositional semantics.  The design of the
  framework is derived from mathematical first principles, in
  particular coalgebraic coinduction and a standard relational model
  of stateful computation.  The abstract syntax and semantics
  introduced here constitute the full core of a novel stream
  programming language.
\end{abstract}


\section{Introduction and Related Paradigms}

For many computation problems a static mapping from input to output
values is not sufficient; outputs need to change over time according
to corresponding changes in inputs and/or internal state.
Applications range from reactive systems, embedded in a context with
user input devices, sensors or communication channels, to processors
and generators of time-series data, such as audio signals or dynamic
simulations in numerous branches of science and engineering.

\subsection{Data-Flow Graphs}

A single paradigm has dominated the landscape of available tools for
the declarative construction of such computational systems, and is
apparently greatly favoured by domain experts without formal training
in programming or software engineering: the visual description in
terms of \emph{data-flow graphs}, made up of computational component
``boxes'' (operations) connected by ``wires'' (variables).  A
characteristic feature is that computations intended to act on whole
streams are written \emph{as if} they were to act on one element at a
time.  Memory of previous elements is retained semi-implicitly by
\emph{delay} components (either boxes or wires) which, unlike other
components, do not denote instantaneous data-flow.

\begin{figure}
  \centering
  \begin{pgfpicture}{-0.5cm}{-0.5cm}{11.5cm}{5.75cm}
    \begin{sigbound}{\pgfxy(0.25,-0.25)}{\pgfxy(10.5,6)}
      \sigport{w}{\pgfxy(0,2.75)}{x}
      \sigline{\pgfxy(0,2.75)}{\pgfxy(9,2.75)}
      \sigbox{\pgfxy(1,4.5)}{\pgfxy(1,1)}{\delta}
      \pgfputat{\pgfxy(3.5,5)}{\pgfbox[center,center]{$\cdots$}}
      \sigbox{\pgfxy(5,4.5)}{\pgfxy(1,1)}{\delta}
      \sigline*{\pgfxy(0.5,2.75)}[\pgfxy(0.5,5)]{\pgfxy(1,5)}
      \sigline{\pgfxy(2,5)}{\pgfxy(3,5)}
      \sigline{\pgfxy(4,5)}{\pgfxy(5,5)}
      \sigline{\pgfxy(6,5)}[\pgfxy(6.5,5)][\pgfxy(6.5,4.25)]{\pgfxy(7,4.25)}
      \sigline*{\pgfxy(4.5,5)}[\pgfxy(4.5,4)]{\pgfxy(7,4)}
      \sigline*{\pgfxy(2.5,5)}[\pgfxy(2.5,3.5)]{\pgfxy(7,3.5)}
      \sigbox{\pgfxy(7,3.25)}{\pgfxy(1,1.25)}{\langle\theta\rvert}
      \sigline{\pgfxy(8,3.75)}[\pgfxy(8.5,3.75)][\pgfxy(8.5,3)]{\pgfxy(9,3)}
      \sigbox{\pgfxy(1,0)}{\pgfxy(1,1)}{\delta}
      \pgfputat{\pgfxy(3.5,0.5)}{\pgfbox[center,center]{$\cdots$}}
      \sigbox{\pgfxy(5,0)}{\pgfxy(1,1)}{\delta}
      \sigbox{\pgfxy(7,1)}{\pgfxy(1,1.25)}{\langle\phi\rvert}
      \sigline{\pgfxy(8,1.75)}[\pgfxy(8.5,1.75)][\pgfxy(8.5,2.5)]{\pgfxy(9,2.5)}
      \sigline*{\pgfxy(10.5,2.75)}[\pgfxy(10.5,0.5)]{\pgfxy(6,0.5)}
      \sigline{\pgfxy(5,0.5)}{\pgfxy(4,0.5)}
      \sigline{\pgfxy(3,0.5)}{\pgfxy(2,0.5)}
      \sigline*{\pgfxy(4.5,0.5)}[\pgfxy(4.5,1.25)]{\pgfxy(7,1.25)}
      \sigline*{\pgfxy(2.5,0.5)}[\pgfxy(2.5,1.75)]{\pgfxy(7,1.75)}
      \sigline{\pgfxy(1,0.5)}[\pgfxy(0.5,0.5)][\pgfxy(0.5,2)]{\pgfxy(7,2)}
      \sigbox{\pgfxy(9,2.25)}{\pgfxy(1,1)}{+}
      \sigline{\pgfxy(10,2.75)}{\pgfxy(11,2.75)}
      \sigport{e}{\pgfxy(11,2.75)}{y}
    \end{sigbound}
  \end{pgfpicture}
  \begin{equation*}
    \bigl(1 - \textstyle\sum \phi_n B^n\bigr) y = \bigl(1 + \textstyle\sum \theta_n B^n\bigr) x
  \end{equation*}
  \pgfimage[width=5in]{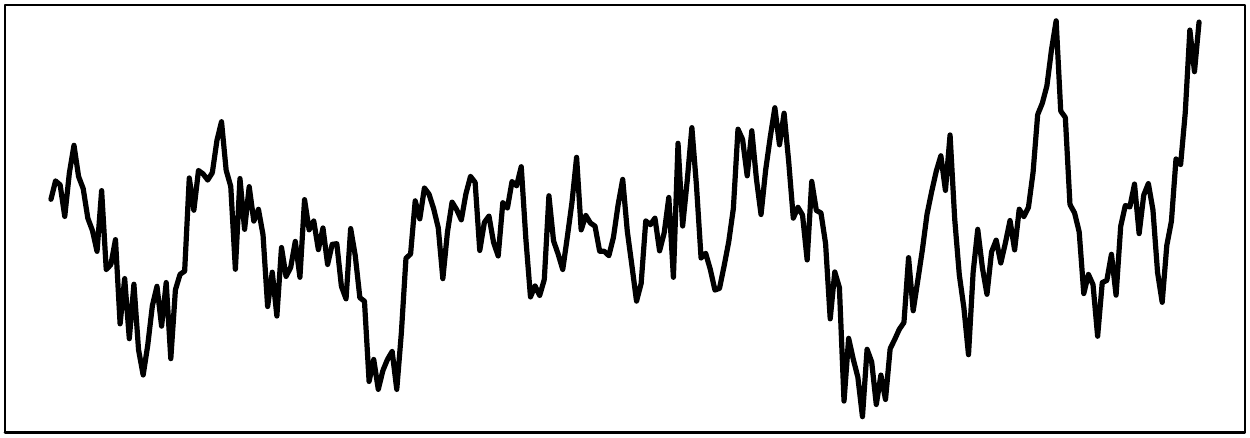}
  \caption{Auto-regressive moving-average (ARMA) model.  {\em Top} --
    data-flow graph: $\delta$ is one-cycle delay; $\langle\phi\rvert$
    is auto-regressive linear form with coefficients $\phi_1, \dots,
    \phi_p$; $\langle\theta\rvert$ is moving-average linear form with
    coefficients $\theta_1, \dots, \theta_q$.  {\em Center} --
    Classical equational presentation in terms of formal power series
    of backshift operator $B$.  {\em Bottom} -- simulated output from
    250 cycles of white noise input, with coefficients
    $\langle\phi\rvert = (0.4, 0.3, 0.2, -0.1)$ and
    $\langle\theta\rvert = (0.3, 0.2, -0.1)$.}
  \label{fig:arma}
\end{figure}

A real-world example is depicted in Figure~\ref{fig:arma}.  It
specifies the basic structure of the auto-regressive--moving-average
(ARMA) model class \cite{Box1970}, which is tremendously influential
in many empirical sciences, for reasons that should be evident from
the shown example model output.  The model is classically presented in
the shown algebraic form, using formal power series of stream
operators.  While that form certainly has its merits for the analytic
treatment of abstract model properties, it is clear that the data-flow
graph of the model, as we draw it here, is more illuminating with
respect to concrete algorithmic details.

The data-flow graph approach, in spite of its intuitive appeal, is
afflicted with several annoying weaknesses of expressivity:
\begin{enumerate}
\item Regular, inverse tree-shaped data flow is far more clumsy than
  in term notations, in particular when noncommutative operators are
  concerned.
\item There is no canonical concept for control issues, such as
  branching, mode transition, self-con\-figura\-tion, initialization.
  In terms of data structures, \emph{product} shape is well-supported,
  whereas \emph{coproduct} shape is not.
  \begin{NEWENV2}
    It is no coincidence that practical systems often provide a
    distinct, not fully integrated language layer for this purpose;
    for instance see the Simulink--Stateflow pair~\cite{simulink}.
  \end{NEWENV2}
\item The view is heavily biased towards \emph{numeric} data and
  operations; their \emph{symbolic} counterparts, essential for
  complex and high-level applications, are treated quite poorly.
\item Static type discipline is rough or entirely missing.  A
  fortiori, software engineering issues, such as \emph{totality} of
  system functions or specification \emph{refinement} relationships,
  cannot be reasoned about.
\end{enumerate}
\begin{NEWENV}
  To compound these problems, reference implementations on all levels,
  from artistic tools \cite{SupercolliderHome,SupercolliderBook} to
  industry-standard modeling frameworks \cite{simulink}, are notorious
  for their awkward behavior, caused by the conflated treatment of
  imperative and declarative features.\footnote{For instance, see a
    complex blog discussion on parameter initialization in Simulink at
    \cite{simulink-params}.}
\end{NEWENV}

\begin{NEWENV}
  Some of these problems are slightly improved by switching to a
  textual representation of data-flow graphs, such as in the Lustre
  core language~\cite{Caspi1987} of the graphical Scade
  environment~\cite{scade}.  Others are apparently consequences of the
  typical low-level machine paradigm underlying the data-flow graph
  approach.
\end{NEWENV}

\subsection{Functional (Reactive) Programming}

Functional programming eliminates many of the shortcomings discussed
above directly, by virtue of its elementary features, such as sound
computational purity, static typing, algebraic data types and pattern
matching.

\newcommand\hanode[4]{
  \begin{pgfscope}
    \color{bg}
    \pgfcircle[fill]{#2}{1.2cm}
  \end{pgfscope}
  \color{fg}
  \pgfnodecircle{#1}[stroke]{#2}{1.2cm}
  \pgfputat{\pgfrelative{\pgfnodecenter{#1}}{\pgfxy(0,0.6)}}{\pgfbox[center,base]{\bfseries\itshape #3}}
  \pgfputat{\pgfrelative{\pgfnodecenter{#1}}{\pgfxy(0,-0.3)}}{\pgfbox[center,center]{\ensuremath{\begin{array}{c}#4\end{array}}}}
}

\begin{figure}
  \begin{center}
  \begin{pgfpicture}{0cm}{0.75cm}{12cm}{7.25cm}
    \pgfsetlinewidth{\sigthickness}
    \pgfsetendarrow{\pgfarrowtriangle{2\sigthickness}}
    \pgfnodesetsepend{3\sigthickness}
    \pgfnodesetsepstart{1.5\sigthickness}
    \colorlet{bg}{black!10}
    \colorlet{fg}{black}
    \hanode{A}{\pgfxy(9,6)}{Attack}{\dot x = +a(x) \\ x < 1 \\ g = 1}
    \hanode{D}{\pgfxy(9,2)}{Decay}{\dot x = -d(x) \\ x > s \\ g = 1}
    \hanode{S}{\pgfxy(3,2)}{Sustain}{\dot x = 0 \\ g = 1}
    \hanode{R}{\pgfxy(3,6)}{Release}{\dot x = -r(x) \\ x \geq 0 \\ g = 0}
    \pgfnodecircle{init}[fill]{\pgfxy(0.5,6)}{1mm}
    \pgfnodeconnline{init}{R}
    \pgfnodeconncurve{R}{A}{15}{165}{1cm}{1cm}
    \pgfnodeconnline{A}{D}
    \pgfnodeconnline{D}{S}
    \pgfnodeconnline{S}{R}
    \pgfnodeconncurve{A}{R}{195}{345}{1cm}{1cm}
    \pgfnodeconnline{D}{R}
    \pgfnodelabel{init}{R}[0.5][2mm]{\pgfbox[center,base]{$x = 0$}}
    \pgfnodelabel{R}{A}[0.5][7mm]{\pgfbox[center,base]{$g = 1$}}
    \pgfnodelabel{A}{R}[0.5][7mm]{\pgfbox[center,top]{$g = 0$}}
    \pgfnodelabel{A}{D}[0.5][2mm]{\pgfbox[left,center]{$x \geq 1$}}
    \pgfnodelabel{D}{S}[0.5][2mm]{\pgfbox[center,top]{$x \leq s$}}
    \pgfnodelabel{S}{R}[0.5][2mm]{\pgfbox[right,center]{$g = 0$}}
    \pgfnodelabelrotated{R}{D}[0.6][2mm]{\pgfbox[center,base]{$g = 0$}}
  \end{pgfpicture}
\end{center}
\begin{hscode}\SaveRestoreHook
\column{B}{@{}>{\hspre}l<{\hspost}@{}}%
\column{3}{@{}>{\hspre}l<{\hspost}@{}}%
\column{5}{@{}>{\hspre}l<{\hspost}@{}}%
\column{12}{@{}>{\hspre}l<{\hspost}@{}}%
\column{14}{@{}>{\hspre}l<{\hspost}@{}}%
\column{21}{@{}>{\hspre}l<{\hspost}@{}}%
\column{23}{@{}>{\hspre}l<{\hspost}@{}}%
\column{30}{@{}>{\hspre}l<{\hspost}@{}}%
\column{32}{@{}>{\hspre}l<{\hspost}@{}}%
\column{36}{@{}>{\hspre}l<{\hspost}@{}}%
\column{39}{@{}>{\hspre}l<{\hspost}@{}}%
\column{41}{@{}>{\hspre}l<{\hspost}@{}}%
\column{50}{@{}>{\hspre}c<{\hspost}@{}}%
\column{50E}{@{}l@{}}%
\column{53}{@{}>{\hspre}l<{\hspost}@{}}%
\column{64}{@{}>{\hspre}c<{\hspost}@{}}%
\column{64E}{@{}l@{}}%
\column{68}{@{}>{\hspre}l<{\hspost}@{}}%
\column{75}{@{}>{\hspre}l<{\hspost}@{}}%
\column{E}{@{}>{\hspre}l<{\hspost}@{}}%
\>[3]{}\mathbf{data}\;\Conid{State}\mathrel{=}\Conid{Attack}\mid \Conid{Decay}\mid \Conid{Sustain}\mid \Conid{Release}{}\<[75]%
\>[75]{}\mbox{\onelinecomment  State space}{}\<[E]%
\\[\blanklineskip]%
\>[3]{}\Varid{adsr}\mathbin{::}(\Conid{Num}\;\Varid{a},\Conid{Ord}\;\Varid{a})\Rightarrow {}\<[30]%
\>[30]{}(\Varid{a}\to \Varid{a})\to (\Varid{a}\to \Varid{a})\to \Varid{a}\to (\Varid{a}\to \Varid{a})\to {}\<[75]%
\>[75]{}\mbox{\onelinecomment  a, d, s, r}{}\<[E]%
\\
\>[30]{}[\mskip1.5mu \Conid{Bool}\mskip1.5mu]\to [\mskip1.5mu \Varid{a}\mskip1.5mu]{}\<[75]%
\>[75]{}\mbox{\onelinecomment  gate, out}{}\<[E]%
\\
\>[3]{}\Varid{adsr}\;\Varid{a\char95 rate}\;\Varid{d\char95 rate}\;\Varid{s\char95 level}\;\Varid{r\char95 rate}\mathrel{=}\Varid{loop}\;\Conid{Release}\;\mathrm{0}{}\<[75]%
\>[75]{}\mbox{\onelinecomment  Initialization}{}\<[E]%
\\
\>[3]{}\hsindent{2}{}\<[5]%
\>[5]{}\mathbf{where}\;{}\<[12]%
\>[12]{}\Varid{loop}\;\Varid{state}\;\Varid{out}\;(\Varid{gate}\mathbin{:}\Varid{gates})\mathrel{=}\Varid{out'}\mathbin{:}\Varid{loop}\;\Varid{state'}\;\Varid{out'}\;\Varid{gates}{}\<[75]%
\>[75]{}\mbox{\onelinecomment  Coinduction}{}\<[E]%
\\
\>[12]{}\hsindent{2}{}\<[14]%
\>[14]{}\mathbf{where}\;{}\<[21]%
\>[21]{}\Varid{out'}\mathrel{=}\mathbf{case}\;\Varid{state}\;\mathbf{of}\;{}\<[75]%
\>[75]{}\mbox{\onelinecomment  Output}{}\<[E]%
\\
\>[21]{}\hsindent{2}{}\<[23]%
\>[23]{}\Conid{Attack}{}\<[32]%
\>[32]{}\to {}\<[36]%
\>[36]{}\Varid{min}\;{}\<[41]%
\>[41]{}\mathrm{1}{}\<[50]%
\>[50]{}\mathbin{\$}{}\<[50E]%
\>[53]{}\Varid{out}\mathbin{+}\Varid{a\char95 rate}\;\Varid{out}{}\<[E]%
\\
\>[21]{}\hsindent{2}{}\<[23]%
\>[23]{}\Conid{Decay}{}\<[32]%
\>[32]{}\to {}\<[36]%
\>[36]{}\Varid{max}\;{}\<[41]%
\>[41]{}\Varid{s\char95 level}{}\<[50]%
\>[50]{}\mathbin{\$}{}\<[50E]%
\>[53]{}\Varid{out}\mathbin{-}\Varid{d\char95 rate}\;\Varid{out}{}\<[E]%
\\
\>[21]{}\hsindent{2}{}\<[23]%
\>[23]{}\Conid{Sustain}{}\<[32]%
\>[32]{}\to {}\<[53]%
\>[53]{}\Varid{out}{}\<[E]%
\\
\>[21]{}\hsindent{2}{}\<[23]%
\>[23]{}\Conid{Release}{}\<[32]%
\>[32]{}\to {}\<[36]%
\>[36]{}\Varid{max}\;{}\<[41]%
\>[41]{}\mathrm{0}{}\<[50]%
\>[50]{}\mathbin{\$}{}\<[50E]%
\>[53]{}\Varid{out}\mathbin{-}\Varid{r\char95 rate}\;\Varid{out}{}\<[E]%
\\
\>[21]{}\Varid{state'}\mathrel{=}\mathbf{case}\;\Varid{state}\;\mathbf{of}\;{}\<[75]%
\>[75]{}\mbox{\onelinecomment  Transition}{}\<[E]%
\\
\>[21]{}\hsindent{2}{}\<[23]%
\>[23]{}\Conid{Release}{}\<[32]%
\>[32]{}\mid {}\<[39]%
\>[39]{}\Varid{gate}{}\<[64]%
\>[64]{}\to {}\<[64E]%
\>[68]{}\Conid{Attack}{}\<[E]%
\\
\>[21]{}\hsindent{2}{}\<[23]%
\>[23]{}\Conid{Attack}{}\<[32]%
\>[32]{}\mid {}\<[39]%
\>[39]{}\Varid{gate}\mathrel{\wedge}\Varid{out'}\geq \mathrm{1}{}\<[64]%
\>[64]{}\to {}\<[64E]%
\>[68]{}\Conid{Decay}{}\<[E]%
\\
\>[21]{}\hsindent{2}{}\<[23]%
\>[23]{}\Conid{Decay}{}\<[32]%
\>[32]{}\mid {}\<[39]%
\>[39]{}\Varid{gate}\mathrel{\wedge}\Varid{out'}\leq \Varid{s\char95 level}{}\<[64]%
\>[64]{}\to {}\<[64E]%
\>[68]{}\Conid{Sustain}{}\<[E]%
\\
\>[21]{}\hsindent{2}{}\<[23]%
\>[23]{}\anonymous {}\<[32]%
\>[32]{}\mid \neg \;{}\<[39]%
\>[39]{}\Varid{gate}{}\<[64]%
\>[64]{}\to {}\<[64E]%
\>[68]{}\Conid{Release}{}\<[E]%
\\
\>[21]{}\hsindent{2}{}\<[23]%
\>[23]{}\anonymous {}\<[64]%
\>[64]{}\to {}\<[64E]%
\>[68]{}\Varid{state}{}\<[E]%
\ColumnHook
\end{hscode}\resethooks
\begin{center}
  \pgfimage[width=4.5in]{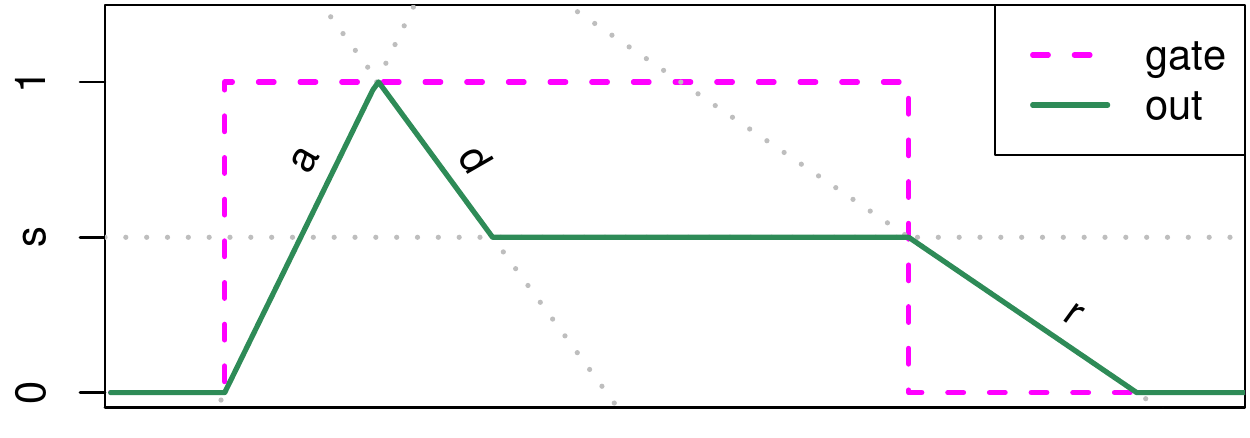}
\end{center}
\caption{Attack--decay--sustain--release (ADSR) device.  {\em Top} --
  Hybrid automaton; implicit constraints: $a(x), d(x),
  r(x) > 0$ except $r(0) = 0$; $0 \leq s \leq 1$.  {\em Center} --
  Discrete-time Haskell implementation; adds guards (before \$) against
  overshooting.  {\em Bottom} -- Example input/output signal pair;
  rates $a, d, r$ constant.}
\label{fig:adsr}
\end{figure}

A real-world example of these principles in action is depicted in
Figure~\ref{fig:adsr}.  It shows the attack--decay--sustain--release
(ADSR) model, which is used in musical synthesis as an amplitude
envelope for the basic soundwave of virtual instruments.  The model
can be described in continous time as a hybrid
automaton~\cite{Henzinger1996}, with four eponymous states.  An output
signal $x$ ranging over the real unit interval $[0,1]$ is computed
from a gate input $g$ over the binary range $\{0, 1\}$.  The shape of
$x$ is controlled by a first-order differential equation per state.
Discrete transitions are triggered either extrinsically by the gate,
or intrinsically by attainment of threshold values.  An implementation
as a Haskell (infinite) list program, involving both discretization in
time and algebraic encoding of the state space, is a fairly
straightforward exercise, as shown.

\begin{NEWENV}
  The paradigm of functional reactive programming (FRP) aims at a
  uniform functional style for very general time-varying data
  (signals), no matter whether change is continuous, asynchronously
  discrete (event-based) or synchronously discrete (clocked).  Stream
  processing arises as a special case, see for instance
  \cite{Wan2000}.  FRP algorithms are typically formulated in embedded
  domain-specific notations hosted in a general-purpose functional
  language.  A main achievement of FRP is a semantic framework that
  keeps the notoriously difficult enforcement of causality under the
  hood, by means of a carefully selected set of programming
  constructs, prescribed by the theory of
  \emph{arrows}~\cite{Hughes2000}.
\end{NEWENV}

The Haskell ADSR implementation of Figure~\ref{fig:adsr}, for all its
elegance \emph{as code}, is theoretically, pragmatically and
philosophically troubling \emph{as a model}: its \emph{complete}
behavior certainly entails the \emph{intended} behavior, but also a
great number of additional cases, effected by the ubiquitous
possibility of bottom values in the semantic CPOs.  We do not believe
that domain experts are generally willing, able or well-advised to
study domain theory (pardon the pun).  There is hardly ever reason for
practical stream functions to be anything else than plain old
mathematical functions.
\begin{NEWENV2}
  The phenomenon of bottom values is of course a fundamental feature
  from the perspective of Turing-complete host environments for stream
  programming, as well as abstract algebraic calculi that aim for a
  general recursive theory, such as the work of
  \Person{Broy}~\cite{Broy1988,Broy2001}.  But from the perspective of
  a mathematically educated user, whether an engineer, scientific
  modeler, or digital artist, is appears more like a defect.  We
  consider this a likely contributing factor to the generally low
  esteem that computational practitioners have for semantics.
\end{NEWENV2}

\section{Discussion and Outline}

\subsection{Vision}

We approach the problem domain from a different angle, with
simultaneously more modest and more ambitious requirements.
\begin{NEWENV3}
  We construct a formal framework that can span all levels of
  abstraction, from loose specification of stream behavior, via
  concrete programs and their implementation techniques, back to
  symbolic and abstract interpretation.
\end{NEWENV3}

On the pragmatical side, we do not abstract from the structure of time
the way FRP does.  Instead, we assume that the system runs in a
clocked synchronous manner.  That is, all input and output data are
equidistant time series, computations are performed each at a fixed,
known clock rate, and data flow is \emph{instantaneous} unless
\emph{delayed} explicitly.  We allow for feedback loops by delayed
circular data flow, but no other forms of recursion.
\begin{NEWENV}
  Note that this rules out some paradoxes of concurrent programming,
  such as the \emph{merge} anomaly; confer~\cite{Broy1988}.
\end{NEWENV}
We envisage that subsystems operating at different clock rates may
coexist in a single system description, but can be sliced and
considered independently for semantic concerns; inter-rate data-flow
is only through dedicated resampling connectors.
\begin{NEWENV}
  Some clock is nevertheless assumed for every stream; this
  distinguishes our approach from the recent coalgebraic treatment of
  index-manipulating (such as sequential splitting and merging)
  operations on streams~\cite{Niqui2013}.
\end{NEWENV}

On the theoretical side, we assume a ``set-theoretical''
framework.\footnote{
  \begin{NEWENV3}
    As opposed to ``domain-theoretical''---we use the term
    ``set-theoretical'' elliptically, as one would in typed
    $\lambda$-calculi.
  \end{NEWENV3}}  Functions realized by components are total, and
observable data never contain bottom values,
confer~\cite{Telford1997}.  We require semantics to be fully
compositional, such that any syntactical fragment has a meaning in the
same domain as the whole.  We also require semantics to support
logical hierarchies of abstraction and refinement of system
descriptions.

On the technical side, we strive for operational compatibility with
existing back-end technology, both in software, such as numeric and
signal-processing algorithm libraries, and in hardware, such as DSP or
FPGA chips.  Note that the actual implementation of the semantic
framework discussed here on such back-ends is future work.

\subsection{Mathematical Tools}

We employ the following tools in the construction of our semantical framework:
\begin{itemize}
\item \Person{Turner}'s notion of total functional
  programming~\cite{Turner2004} and the corresponding dual modes of
  inductive reasoning for data, and coinductive reasoning for
  codata~\cite{Telford1997};
\item \Person{Rutten}'s categorial universal
  coalgebra~\cite{Rutten2000} as a highly abstract foundation of
  coinduction;
\item the relational model of the Z notation~\cite{Spivey1988} for
  stateful and possibly nondeterministic computation;
\item \Person{Pardo}'s monadic coinduction for the unification of the
  preceding two;
\item Static single-assignment (SSA) form~\cite{Cytron1991} for the
  declarative core representation of programs.
\end{itemize}

\begin{NEWENV3}
  While this may seem a baroque and eclectic programme at first sight,
  all tools are considered standard and fundamental in the respective,
  disparate fields, and in each case we extract the minimum of
  features required to solve the posed problem.
\end{NEWENV3}

\subsection{This Paper}

It is not the purpose of this paper to belittle or criticize the
efforts of FRP.  Instead, we pursue a dialectical programme, deriving
an antithesis from distinct and complementary real-world requirements.
Since our approach is still very much work in progress, we shall have
to leave attempts at a possible synthesis of FRP and our own approach
for later.  Hence we do not currently draw a more elaborate technical
comparison either.

The goal of the present paper is to demonstrate how the semantic
ingredients work together, from the perspective of fundamental
programming language design.  It should not be understood as a
recommendation on front-end notation and programming style, or as the
description of an implementation strategy or actual compiler.  The
ongoing work on those aspects shall be discussed in due time in
dedicated companion papers.
\begin{NEWENV}
  However, the basic design and paradigm of a future, complete and
  practically useful programming language is specified here as
  completely as possible within reasonable space limits.
\end{NEWENV}

\begin{NEWENV}
  The derivation of our semantic framework from the purely
  mathematical perspective shall be described in greater detail,
  including proofs, in a forthcoming companion paper.  The present
  paper is organized as follows: Section~\ref{math} introduces the
  mathematical ingredients of semantic objects and their properties.
  Section~\ref{strategy} outlines the idea of semantics assignment to
  compositional descriptions of stream computations, in terms of a
  novel extension to the well-known SSA form of intermediate
  languages.  Section~\ref{core}, the focus and main contribution of
  the present paper, deals with the syntax of a core language and
  corresponding inductive semantic translation and analysis rules.
\end{NEWENV}

\section{Mathematical Foundations}
\label{math}

\subsection{Streams, Set-Theoretically}

On behalf of domain experts with classical mathematical training, we
agree wholeheartedly with \Person{Turner}~\cite{Turner2004}:
\begin{quote}
  \itshape The driving idea of functional programming is to make
  programming more closely related to
  mathematics. \textnormal{[\dots]} The existing model of functional
  programming, although elegant and powerful, is compromised to a
  greater extent than is commonly recognised by the presence of
  partial functions.
\end{quote}

\Person{Turner} proposes to distinguish sharply between \emph{data}
and \emph{codata} with associated inductive and coinductive models of
computation, respectively.  In this terminology, streams are the
paradigmatic example of a codata structure~\cite{Telford1997}.

A set-theoretic account of streams (that is data of the infinite
sequence form $A^\omega$ for some set $A$) and stream functions (that
is, total functions of type $A^\omega \to B^\omega$ which obey some form of
causality and can be evaluated online) is most elegantly given in
terms of categorial coalgebra; confer the seminal work of
Rutten~\cite{Rutten2000}.  The resulting theory is rather simpler and
more regular than its counterparts over CPOs for partial and/or
non-strict functional programming.

The endofunctor $\Stream{A}(X) = A \times X$, or $\Stream{A} = A
\times {-}$ for short, on the category $\mathbf{Set}$ has coalgebras
of the form $(S, \langle h, t \rangle)$ with
\begin{NEWENV2}
  \emph{carrier} or \emph{state space} $S$ and \emph{operations}
\end{NEWENV2}
$h : S \to A$ and $t : S
\to S$.  The coalgebra $(A^\omega, \langle \mathit{head}_A,
\mathit{tail}_A\rangle)$, with $\mathit{head}_A(a) = a_0$ and
$\mathit{tail}_A(a)_n = a_{n+1}$ is final.  That is,
\begin{itemize}
\item the operation $\langle \mathit{head}_A, \mathit{tail}_A \rangle$
  is bijective, with inverse $\mathit{cons}_A : A \times A^\omega \to
  A^\omega$;
\item for every coalgebra $(S, \langle h, t \rangle)$ there is a
  unique \emph{coiteration} homomorphism (or \emph{anamorphism})
  $\ana{h, t}_A : S \to A^\omega$, given by $\ana{h, t}_A(x)_n =
  h\bigl(\underbrace{t(\dots (t}_n(x)) \dots)\bigr)$.
\end{itemize}
The access operations $\mathit{head}$ and $\mathit{tail}$ generalize
inductively to $\mathit{take}_A(n) : A^\omega \to A^n$ and
$\mathit{drop}_A(n) : A^\omega \to A^\omega$ in the obvious way, with
$\mathit{head} = \mathit{take}(1)$ and $\mathit{tail} =
\mathit{drop}(1)$.  Note that, unlike for the familiar Haskell list
counterparts, there are no spurious corner cases; the range of
$\mathit{take}(n)$ is exactly $A^n$.
\begin{NEWENV}
  Thus we obtain an \emph{unrolling} similarity transform $^{(n)}$ for
  $n > 0$ on stream functions:
  \begin{align*}
    f : A^\omega \to B^\omega \implies f^{(n)} : (A^n)^\omega \to
    (B^n)^\omega && f^{(n)} &= \mathit{chop}_B(n) \circ f \circ
    \mathit{chop}_A^{-1}
    \\
    && \text{where}\quad \mathit{chop}_A(n) &= \ana{\mathit{take}_A(n),
      \mathit{drop}_A(n)}_{A^n}
  \end{align*}
\end{NEWENV}

In the same vein, the operation $\mathit{zip}_{AB} : A^\omega \times
B^\omega \to (A \times B)^\omega$ is a natural bijection.  This is
quite useful, because it allows us to reduce stream functions with
multiple arguments and/or results to the unary base case.
\begin{NEWENV}
  In particular, it gives rise to a tensor-product-like operation
  $\otimes$ on stream functions, by a similarity transform:
  \begin{align*}
    \begin{aligned}
      f &: A^\omega \to B^\omega
      \\
      g &: C^\omega \to D^\omega
    \end{aligned} \implies  f \otimes g : (A \times C)^\omega \to (B \times D)^\omega
    && f \otimes g = \mathit{zip}_{BD} \circ (f \times g) \circ
    \mathit{zip}_{AC}^{-1}
  \end{align*}
\end{NEWENV}

Causal stream functions can be defined inductively in the classical
way, as precisely the functions $f : A^\omega \to B^\omega$ such that
$\mathit{take}_A(n)(x) = \mathit{take}_A(n)(y)$ implies
$\mathit{take}_B(n)\bigl(f(x)\bigr) =
\mathit{take}_B(n)\bigl(f(y)\bigr)$ for all $n$.  We write $A^\omega
\causto B^\omega$ for the set of all such functions.  The causality
property implies the existence of lifted operations
$\mathit{take}_{AB}(n) : (A^\omega \causto B^\omega) \to A^n \to B^n$
and $\mathit{drop}_{AB}(n) : (A^\omega \causto B^\omega) \to A^n \to
A^\omega \causto B^\omega$, such that
\begin{align*}
  \mathit{take}_{AB}(n)(f) \circ \mathit{take}_A(n) &=
  \mathit{take}_B(n) \circ f \\
  \mathit{uncurry}\bigl(\mathit{drop}_{AB}(n)(f)\bigr) \circ \langle
  \mathit{take}_A(n), \mathit{drop}_A(n) \rangle &= \mathit{drop}_B(n)
  \circ f
\end{align*}
for all $n$.  We abbreviate $\mathit{head}_{AB} =
\mathit{take}_{AB}(1)$ and $\mathit{tail}_{AB} =
\mathit{drop}_{AB}(1)$ as obvious.

The endofunctor $\Trans[AB] = \mathrm{Hom}(A, {-}) \circ \Stream{B} =
(A \to B \times {-})$ on the category $\mathbf{Set}$ has coalgebras of
the form $(S, f : S \to A \to B \times S)$, where operations $f$ are
in one-to-one correspondence with pairs $h : S \to A \to B$ and $t : S
\to A \to S$, namely by $\mathit{uncurry}(f) = \langle
\mathit{uncurry}(h), \mathit{uncurry}(t) \rangle$, which we abbreviate
to $f = \langle\!\langle h, t \rangle\!\rangle$.  The coalgebra
$(A^\omega \causto B^\omega, \langle\!\langle \mathit{head}_{AB},
\mathit{tail}_{AB} \rangle\!\rangle)$ is final.  That is,
\begin{itemize}
\item the operation $\langle\!\langle \mathit{head}_{AB},
  \mathit{tail}_{AB} \rangle\!\rangle$ is bijective, with inverse
  $\mathit{cons}_{AB} : \bigl(A \to B \times (A^\omega \causto
  B^\omega)\bigr) \to A^\omega \causto B^\omega$;
\item for every coalgebra $(S, \langle\!\langle h, t
  \rangle\!\rangle)$ there is a unique \emph{coiteration} homomorphism
  $\ana{h, t}_A : S \to A^\omega$, given by $\ana{h,
    t}_{AB}(x)(a)_n =
  h\bigl(\underbrace{t(\dots(t}_n(x)(a_0))\dots)(a_{n-1})\bigr)(a_n)$.
\end{itemize}

Note that the isomorphism $A^\omega \causto B^\omega \cong A \to B
\times (A^\omega \causto B^\omega)$ also plays a central role in
continuation-based implementations of FRP, see~\cite{Nilsson2002},
which arrive at the same structure by different arguments.

\subsection{Stateful Relations}
\label{rel}

As we have just seen, causal stream functions can be represented
coalgebraically by a hidden state space $S$ and an operation $f$ with
type $\mathit{uncurry}(f) : S \times A \to B \times S$, which takes
pre-states and inputs to post-states and outputs.  In stream
programming, we generally wish to keep state under the hood.  A
traditional functional solution in the spirit of
\Person{Moggi}~\cite{Moggi1991} would be to massage the operation to
obtain type $A \to S \to B \times S$, which can then be studied in the
Kleisli category $\mathbf{Kl}(M)$ of a state monad $M = S \to {-}
\times S$.  However, that approach is orthogonal to the coalgebraic
one, seeing as they fix complementary type parameters.

Hence we pursue a different direction, by noting the similarity to the
representation of state-based systems in the formal notation
Z~\cite{Spivey1988}, where pre-state, input, output and post-state
variables are distinguished by decorating suffixes, namely nothing,
$?$, $!$ and $'$, respectively.  There, state is handled explicitly,
which we shall emulate by inference en route from a front-end notation
to the core calculus of our proposed framework.  How the algebra of Z
schemas relates to the theories of monads and arrows prevailing in FRP
has, to our knowledge, not been established precisely.  The former is
certainly less abstract, and founded more on pragmatic utility and
trusted first-order logic than axiomatic elegance.  However, a
framework of quaternary stateful relations can be put to good use in
explicating the fundamentals of stream programming semantics, as we
shall presently outline.

\begin{figure}
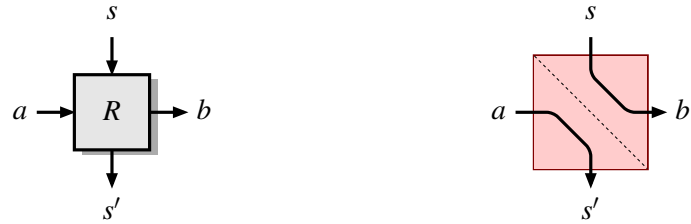

  \centering
  \hbox{}\hfill
  \begin{pgfpicture}{-1cm}{-1cm}{2cm}{2cm}
    \sigbox{\pgfxy(0,0)}{\pgfxy(1,1)}{R}
    \sigline{\pgfxy(-0.5,0.5)}{\pgfxy(0,0.5)}
    \sigport{w}{\pgfxy(-0.5,0.5)}{a}
    \sigline{\pgfxy(0.5,1.5)}{\pgfxy(0.5,1)}
    \sigport{n}{\pgfxy(0.5,1.5)}{s}
    \sigline{\pgfxy(1,0.5)}{\pgfxy(1.5,0.5)}
    \sigport{e}{\pgfxy(1.5,0.5)}{b}
    \sigline{\pgfxy(0.5,0)}{\pgfxy(0.5,-0.5)}
    \sigport{s}{\pgfxy(0.5,-0.5)}{s'}
  \end{pgfpicture}
  \hfill
  \begin{pgfpicture}{-1cm}{-1cm}{2cm}{2cm}
    \begin{sigbound}{\pgfxy(-0.25,-0.25)}{\pgfxy(1.5,1.5)}
      \sigline{\pgfxy(-0.5,0.5)}[\pgfxy(0,0.5)][\pgfxy(0.5,0)]{\pgfxy(0.5,-0.5)}
      \sigline{\pgfxy(0.5,1.5)}[\pgfxy(0.5,1)][\pgfxy(1,0.5)]{\pgfxy(1.5,0.5)}
      \sigport{w}{\pgfxy(-0.5,0.5)}{a}
      \sigport{n}{\pgfxy(0.5,1.5)}{s}
      \sigport{e}{\pgfxy(1.5,0.5)}{b}
      \sigport{s}{\pgfxy(0.5,-0.5)}{s'}
      \pgfsetdash{{\sigthickness}{\sigthickness}}{0pt}
      \pgfxyline(-0.25,1.25)(1.25,-0.25)
    \end{sigbound}
  \end{pgfpicture}
  \hfill\hbox{}
  \caption{Stateful relations: {\em Left} -- general graphical layout.
    {\em Right} -- single-step delay component $\delta$.}
  \label{fig:rel}
\end{figure}

The algebra of stateful relations is best explained visually.  Each
relation can be depicted as a quadrilateral box, where the sides
correspond to variable roles by convention: I/O and state transition
are layed out left-to-right and top-to-bottom, respectively; see
Figure~\ref{fig:rel}.  The four sides can in general each have product
structure, corresponding to zero or more variables, and be connected
in arbitrary ways.  For practical purposes, however, it is more
desirable to consider building blocks with more disciplined behavior.
For instance, stateless operations (the vast majority in many stream
function definitions) have zero state variables and hence unit state.
The most fundamental nontrivial stateful relation is the single-step
delay, which transfers input to post-state and pre-state to output,
respectively; see also Figure~\ref{fig:rel}.

There are two sensible ways to encode relations,
\begin{enumerate}
\item as tuple subsets of $S \times A \times B \times S$, thereby
  inheriting the Boolean lattice structure of entailment that gives a
  hierarchy of behavioral descriptions, essential both a priori for
  nondeterministic specifications and refinement, and a posteriori for
  program abstraction and slicing;
\item equivalently and more conveniently for the following
  computational discussions, as set-valued maps of type $S \times A
  \to M(B \times S)$, where $M$ is some variant of the powerset monad.
  These maps can be studied in the respective Kleisli category
  $\mathbf{Kl}(M)$, which yields general relations for the
  unrestricted powerset $M = \Power$, and left-total/functional
  relations for the restrictions $M = \Power*/\PowerI$ to
  nonempty/singular subsets, respectively.  Note that the latter is
  merely a ``disguised'' equivalent of the identity monad; it helps to
  address functions as particular relations in the common relational
  language.  For the differentiated use of all three candidates, see
  section~\ref{compositional} below.
\end{enumerate}

\begin{figure}
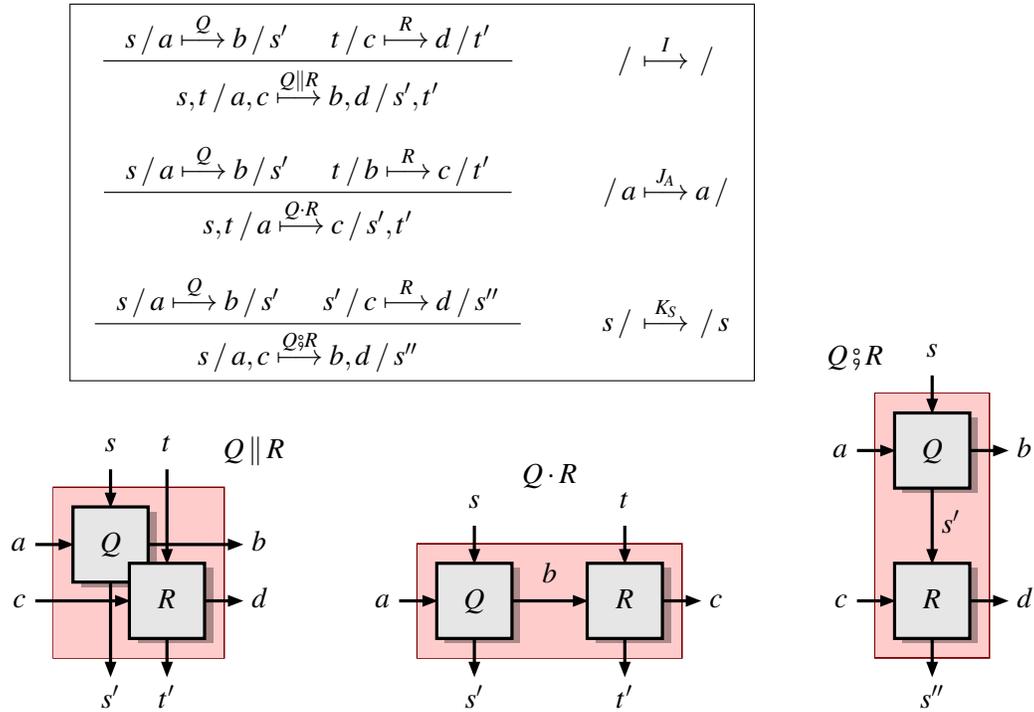

  \centering
\begin{equation*}
  \boxed{
  \begin{array}{c@{\hspace{1cm}}c}
    \inference{s \insep a \stackrel{Q}\longmapsto b \outsep s' & t \insep c \stackrel{R}\longmapsto d \outsep t'}{s, t \insep a, c \stackrel{Q \mathbin\| R}\longmapsto b, d \outsep s', t'}
    & {} \insep {} \stackrel{I}\longmapsto {} \outsep {}
    \\[5ex]
    \inference{s \insep a \stackrel{Q}\longmapsto b \outsep s' & t \insep b \stackrel{R}\longmapsto c \outsep t'}{s, t \insep a \stackrel{Q \cdot R}\longmapsto c \outsep s', t'}
    & {} \insep a \stackrel{J\!_A}\longmapsto a \outsep {}
    \\[5ex]
    \inference{s \insep a \stackrel{Q}\longmapsto b \outsep s' & s' \insep c \stackrel{R}\longmapsto d \outsep s''}{s \insep a, c \stackrel{Q \fatsemi R}\longmapsto b, d \outsep s''}
    & s \insep {} \stackrel{K\!_S}\longmapsto {} \outsep s
  \end{array}}
  \hspace{3cm}
\end{equation*}
\\[-1cm]
\hfill
  \begin{pgfpicture}{-1cm}{-1.75cm}{2.75cm}{2cm}
    \begin{sigbound}{\pgfxy(-0.25,-1)}{\pgfxy(2.25,2.25)}
      \pgfputat{\pgfxy(2,1.75)}{\pgfbox[left,center]{$Q \mathbin\| R$}}
      \sigbox{\pgfxy(0,0)}{\pgfxy(1,1)}{Q}
      \sigbox{\pgfxy(0.75,-0.75)}{\pgfxy(1,1)}{R}
      \sigline{\pgfxy(-0.5,0.5)}{\pgfxy(0,0.5)}
      \sigport{w}{\pgfxy(-0.5,0.5)}{a}
      \sigline{\pgfxy(0.5,1.5)}{\pgfxy(0.5,1)}
      \sigport{n}{\pgfxy(0.5,1.5)}{s}
      \sigline{\pgfxy(1,0.5)}{\pgfxy(2.25,0.5)}
      \sigport{e}{\pgfxy(2.25,0.5)}{b}
      \sigline{\pgfxy(0.5,0)}{\pgfxy(0.5,-1.25)}
      \sigport{s}{\pgfxy(0.5,-1.25)}{s'}
      \sigline{\pgfxy(-0.5,-0.25)}{\pgfxy(0.75,-0.25)}
      \sigport{w}{\pgfxy(-0.5,-0.25)}{c}
      \sigline{\pgfxy(1.25,1.5)}{\pgfxy(1.25,0.25)}
      \sigport{n}{\pgfxy(1.25,1.5)}{t}
      \sigline{\pgfxy(1.75,-0.25)}{\pgfxy(2.25,-0.25)}
      \sigport{e}{\pgfxy(2.25,-0.25)}{d}
      \sigline{\pgfxy(1.25,-0.75)}{\pgfxy(1.25,-1.25)}
      \sigport{s}{\pgfxy(1.25,-1.25)}{t'}
    \end{sigbound}
  \end{pgfpicture}
  \hfill
  \begin{pgfpicture}{-1cm}{-1cm}{4cm}{2cm}
    \begin{sigbound}{\pgfxy(-0.25,-0.25)}{\pgfxy(3.5,1.5)}
      \pgfputat{\pgfxy(1.5,2)}{\pgfbox[center,bottom]{$Q \cdot R$}}
      \sigbox{\pgfxy(0,0)}{\pgfxy(1,1)}{Q}
      \sigbox{\pgfxy(2,0)}{\pgfxy(1,1)}{R}
      \sigline{\pgfxy(-0.5,0.5)}{\pgfxy(0,0.5)}
      \sigport{w}{\pgfxy(-0.5,0.5)}{a}
      \sigline{\pgfxy(0.5,1.5)}{\pgfxy(0.5,1)}
      \sigport{n}{\pgfxy(0.5,1.5)}{s}
      \sigline{\pgfxy(1,0.5)}{\pgfxy(2,0.5)}
      \sigport{n}{\pgfxy(1.5,0.5)}{b}
      \sigline{\pgfxy(0.5,0)}{\pgfxy(0.5,-0.5)}
      \sigport{s}{\pgfxy(0.5,-0.5)}{s'}
      \sigline{\pgfxy(2.5,1.5)}{\pgfxy(2.5,1)}
      \sigport{n}{\pgfxy(2.5,1.5)}{t}
      \sigline{\pgfxy(3,0.5)}{\pgfxy(3.5,0.5)}
      \sigport{e}{\pgfxy(3.5,0.5)}{c}
      \sigline{\pgfxy(2.5,0)}{\pgfxy(2.5,-0.5)}
      \sigport{s}{\pgfxy(2.5,-0.5)}{t'}
    \end{sigbound}
  \end{pgfpicture}  
  \hfill
  \begin{pgfpicture}{-1cm}{-3cm}{2cm}{2cm}
    \begin{sigbound}{\pgfxy(-0.25,-2.25)}{\pgfxy(1.5,3.5)}
      \pgfputat{\pgfxy(-0.9,1.9)}{\pgfbox[left,top]{$Q \fatsemi R$}}
      \sigbox{\pgfxy(0,0)}{\pgfxy(1,1)}{Q}
      \sigbox{\pgfxy(0,-2)}{\pgfxy(1,1)}{R}
      \sigline{\pgfxy(-0.5,0.5)}{\pgfxy(0,0.5)}
      \sigport{w}{\pgfxy(-0.5,0.5)}{a}
      \sigline{\pgfxy(0.5,1.5)}{\pgfxy(0.5,1)}
      \sigport{n}{\pgfxy(0.5,1.5)}{s}
      \sigline{\pgfxy(1,0.5)}{\pgfxy(1.5,0.5)}
      \sigport{e}{\pgfxy(1.5,0.5)}{b}
      \sigline{\pgfxy(0.5,0)}{\pgfxy(0.5,-1)}
      \sigport{e}{\pgfxy(0.5,-0.5)}{s'}
      \sigline{\pgfxy(-0.5,-1.5)}{\pgfxy(0,-1.5)}
      \sigport{w}{\pgfxy(-0.5,-1.5)}{c}
      \sigline{\pgfxy(1,-1.5)}{\pgfxy(1.5,-1.5)}
      \sigport{e}{\pgfxy(1.5,-1.5)}{d}
      \sigline{\pgfxy(0.5,-2)}{\pgfxy(0.5,-2.5)}
      \sigport{s}{\pgfxy(0.5,-2.5)}{s''}
    \end{sigbound}
  \end{pgfpicture}  
  \hfill\hbox{}
  \caption{Stateful relational compositions, with inductive definitions and neutral elements}
  \label{fig:compose}
\end{figure}
Either way, we use the graphically evocative notation $s \mathrel/ a
\mapsto b \mathrel/ s'$ for pointwise reasoning.  It comes with two
mnemonic aids:
\begin{enumerate}
\item $s$ (top) above $a$ (left) goes to $b$ (right) above $s'$ (bottom);
\item the I/O relation is $a \mapsto b$, embedded in the state context
  $s \mathrel/ \dots \mathrel/ s'$.
\end{enumerate}

In the visual data-flow representation, the boxes depicting relations
are truly three-dimensional objects.  There are three distinct
composition operations on stateful relations that make intuitive
sense.  Each is associative and admits neutral elements, thus giving a
monoidal structure (up to type compatibility).  See
Fig.~\ref{fig:compose}.  The rules become more evidently convincing
when compared to the visual form.  Note that ordinary relational
composition, simultaneously affecting both state and I/O, has no place
in this calculus.

With respect to stream computations, the state axis of our relations
is special because, although flat on the element level, a feedback
loop is implied on the stream level, with post-state of each step
becoming pre-state of the following.  Informally, an element-level
relation $R$ induces a stream-level function $g$ as
\begin{equation*}
  \text{``}\enspace g = \lim_{n \to \omega} \;\underbrace{R \fatsemi \dots \fatsemi R}_n \enspace\text{''}
\end{equation*}
There is no obvious way, such as a topology, to take that limit
literally.  We shall demonstrate in the following subsection that the
coalgebraic approach can be extended to solve the corresponding
fixpoint equation $g \cong R \fatsemi g$ in a canonical way.

\subsection{Relational Coinduction}
\label{rel-ana}

The framework of monadic coinduction, although introduced specifically
over categories of CPOs by \Person{Pardo}~\cite{Pardo1998}, carries
over nicely to our stateful relations over $\mathbf{Set}$.  All three
candidate functors $M = \Power, \Power*, \PowerI$ admit the required
structure:
\begin{itemize}
\item They are commutative strong monads with strength $\tau_{X,Y} : X
  \times M(Y) \to M(X \times Y)$.  Note that $\tau_X$ for any fixed
  $X$ is a distributive law of $\Stream{X}$ over $M$.
\item They come with distributive laws for the functors
  $\mathrm{Hom}(X, {-})$ for any $X$, namely
  \begin{NEWENV3}
    the choice function operator
  \end{NEWENV3}
  $\lambda_{X,Y} : \bigl(X \to M(Y)\bigr) \to M(X \to Y)$
  \begin{NEWENV3}
    with $\lambda_{X,Y}(f) = \{ g \mid \operatorname\forall x
    \mathpunct. g(x) \in f(x) \}$.
  \end{NEWENV3}
  Note that
    the case $M = \Power*$ corresponds to the axiom of choice.  This
  law establishes a correspondence of stateful relations with
  $\Trans[A,B]$-coalgebras over the Kleisli category $\mathbf{Kl}(M)$:
  \begin{equation*}
    R : S \times A \to M(B \times S) \implies
    \underbrace{\lambda_{A,B \times S} \circ
      \mathit{curry}(R)}_{f_R} : S \to M(\underbrace{A \to B \times
      S}_{\Trans[A,B](S)})
  \end{equation*}
\item The two distributive laws composed yield a distributive law for
  $\Trans[A,B]$ over $M$ that satisfies suitable coherence laws to
  induce a lifting $\LTrans[A,B]$ of $\Trans[A,B]$ from the category
  $\mathbf{Set}$ to $\mathbf{Kl}(M)$.
  \begin{align*}
    \LTrans[A,B](X) &= \Trans[A,B](X) & f : X \to M(Y) \implies \LTrans[A,B](f) : \LTrans[A,B](X) \to M\bigl(\LTrans[A,B](Y)\bigr)
  \end{align*}
\end{itemize}
Thus prepared, we can specify $M$-monadic $\Trans[A,B]$-coinduction
following \cite{Pardo1998}: To a $\Trans[A,B]$-coalgebra $(S, f)$ over
$\mathbf{Kl}(M)$, and in particular to one arising from a stateful
relation, we associate homomorphisms $g$ such that the following
diagram commutes, where $\_^\star$ denotes Kleisli extension:
\begin{equation*}
  \vcenter{\xymatrix{
      S \ar[d]_{f} \ar[r]^-{g} & M(A^\omega \causto B^\omega) \ar[d]^{M(\mathit{cons}_{AB}^{-1})}
      \\
      M\bigl(\Trans[A,B](S)\bigr) \ar[r]_-{\LTrans[A,B](g)^\star} & M\bigl(\Trans[A,B](A^\omega \causto B^\omega)\bigr)
    }}
\end{equation*}
This diagram does not generally have a unique solution, and
\cite{Pardo1998} discusses the existence and possible choices over the
category $\mathbf{CPO}$ in detail.  We need to choose a canonical
solution $g = \ana{f}^M_{AB}$ over the category $\mathbf{Set}$
instead.  Fortunately, the existence argument carries over from
$\mathbf{CPO}$.  Noting that the right vertical arrow is the
invertible operation of the final $\Trans[A,B]$-coalgebra, we find the
desired fixpoint equation:
\begin{equation*}
  g = \underbrace{M(\mathit{cons}_{AB}) \circ \LTrans[A,B](g)^\star \circ f}_{\Phi(g)}
\end{equation*}

The candidate monad $M = \Power$ carries a complete lattice structure
that can be raised to the $S$-power.  The operator $\Phi$ is easily
seen to be monotone on this structure.  Hence \Person{Tarski}'s
theorem guarantees a nonempty choice of solutions, including two
canonical choices, namely the least and the greatest fixpoint.  As a
general rule, least fixpoints are preferred over $\mathbf{CPO}$ in
general, and for induction over $\mathbf{Set}$, but greatest fixpoints
are preferred for coinduction over $\mathbf{Set}$.  This rule of thumb
is confirmed here:
\begin{itemize}
\item the least fixpoint is the empty relation, a lifting of the fact
  the the initial $\Stream{A}$-algebra is also empty;
\item the more restricted candidate monad $M = \Power*$ has the same upper bounds,
  but generally no lower bounds.
\end{itemize}
The most restricted candidate monad $M = \PowerI$ has unique solutions
by equivalence to the identity monad.  Hence we can uniformly choose
the greatest fixpoint in either case.

In summary, monadic coinduction takes each
(general/left-total/functional) stateful relation canonically to a
(general/left-total/functional) relation between initial states and
possible observable behavior, that is, causal I/O stream functions.
\begin{NEWENV}
  \begin{align*}
    R : S \times A \to M(B \times S) \implies \sem{R}^M = \ana{f_R}^M : S \to M(A^\omega \causto B^\omega)
  \end{align*}
\end{NEWENV}%
\begin{NEWENV2}%
  To our knowledge, this is a novel application of monadic coinduction.
\end{NEWENV2}

\subsection{Compositional Denotational Semantics}
\label{compositional}

Our approach to semantics poses a dialectical problem: On one hand, we
intend to give total stream function semantics to box-like program
components.  On the other hand, we require semantics for program
fragments to be compositional, including individual pattern matching
rules, which are partial by their very essence.  The synthesis that
solves this problem involves a differentiation of several semantic
meaning functions.

Let $\sem[V]{x}$ denote the range of possible values for a variable
$x$.  This can be established by declared or inferred type
information, or simply be the universe of terms in the untyped case.
Note that, in a total functional setting, no undefined value $\bot$ is
included by default.  Instead, we write explicitly $\sem[V_\bot]{x} =
\sem[V]{x} \uplus \{ \bot \}$ for the extension.  Range assignment lifts
to vectors of variables by taking the non-strict Cartesian product:
\begin{align*}
  \sem[V]{x_1, \dots, x_n} &= \sem[V]{x_1} \times \dots \times
  \sem[V]{x_n} &
  \sem[V_\bot]{x_1, \dots, x_n} &= \sem[V_\bot]{x_1} \times \dots \times
  \sem[V_\bot]{x_n}
\end{align*}
It is possible for variables to have unit range $\sem[V]{x} = \{ \top
\}$.  While these have no effect on data flow in a total functional
setting, the extension $\sem[V_\bot]{x} = \{ \top, \bot \}$ serves as
a primitive Boolean type, which we shall exploit for reducing control
flow to data flow.  Hence we call variables \emph{control} variables,
if they have unit range, and \emph{data} variables otherwise.  In the
data-flow graph notation, control wires are indicated by dashed
arrows.  See Figure~\ref{fig:sah}
below for examples.

Then we can give a hierarchy of three element-wise semantics of
(fragments of) stream programs, in terms of stateful relations.  Each
of these uses a different candidate monad; stream-wise semantics can
be obtained by monadic coinduction at each level:
\begin{enumerate}
\item The compositional but partial, \emph{internal} relational
  semantics of a program fragment with interface $s \mathbin/ x \to y
  \mathbin/ s'$, where all four meta-variables stand for vectors of
  object-level variables, is a left-total relation $R : S \times A \to
  \Power*(B \times S)$ with $A = \sem[V_\bot]{x}$, $B =
  \sem[V_\bot]{y}$ and $S = \sem[V_\bot]{s} = \sem[V_\bot]{s'}$.
  \begin{NEWENV}
    Left-totality is appropriate because, in the general case that
    includes both partial fragments and nondeterministic
    specifications, each post-state and output variable can be
    undefined, or take one or more possible values, or both, but
    clearly not neither.  By coinduction we obtain a nonempty set of
    possible stream functions for each initial state:
    \begin{equation*}
      R : S \times A \to \Power*(B \times S) \implies \sem{R}^{\Power*} : S \to \Power*(A^\omega \causto B^\omega)
    \end{equation*}

    Note that the strongly synchronous model of computation which we
    have adopted avoids the well-known paradoxes of nondeterministic
    concurrent computation, such as the \emph{merge} anomaly and the
    Brock--Ackermann anomaly, which could ruin compositionality in a
    more general setting.
  \end{NEWENV}
\item At box boundaries, we additionally consider the \emph{external}
  relational semantics, which are totalized along the I/O axis by
  component-wise exclusion of $\bot$.  Clearly, the result is no
  longer generally left-total; hence we obtain a subrelation $R^\dag :
  S \times A^\dag \to \Power(B^\dag \times S)$ with $A^\dag =
  \sem[V]{x} \subseteq A$ and $B^\dag = \sem[V]{y} \subseteq B$.

  \begin{NEWENV}
    The transformation has two effects: Firstly, an output variable
    can no longer be considered undefined if it could also take some
    value; thus definedness is maximized individually.  Secondly, all
    output variables need to be defined simultaneously for a valid
    output tuple.
  \end{NEWENV}
  
  This is already the desired element-wise semantics for possibly
  nondeterministic specifications of whole components.
  \begin{NEWENV}
    Since $\bot$ is no valid data element in inter-component flow,
    components can communicate regardless of their respective error
    behavior.  This is particularly beneficial in the presence of
    legacy code and/or hybrid hardware--software implementations.  By
    coinduction we obtain a possibly empty set of possible stream functions
    for each initial state:
    \begin{equation*}
      R^\dag : S \times A^\dag \to \Power(B^\dag \times S) \implies \sem{R^\dag}^{\Power} : S \to \Power\bigl((A^\dag)^\omega \causto (B^\dag)^\omega\bigr)
    \end{equation*}
  \end{NEWENV}

  Note that state is \emph{not} totalized for compositionality
  reasons: the internal state variables of a box can be undefined at
  any time, otherwise we could not give semantics to local delayed
  data flow in pattern matching rules.  However, undefined values may
  not leak to outputs.

\item
  \begin{NEWENV}
    At the level of a specification or abstraction of a program, it
    may be permitted for a functional block to mean zero or multiple
    causal stream functions.  At the level of the concrete program, a
    deterministic meaning is required, which may constrain admissible
    pre- and post-states.  This problem amounts to finding an
    acceptable state space $D \subseteq S$, typically by exclusion of
    $\bot$ from selected components, such that the corresponding
    domain-restricted subrelation is functional; namely $s \in D$
    implies $R^\dag(s \mathbin/ x) = \{ y \mathbin/ s')$ with $s' \in
    D$ also.  We write $R^{[D]} : S \times A^\dag \to \PowerI(B \times
    S)$.  By coinduction we obtain a singleton set of possible stream
    functions for each initial state:
    \begin{equation*}
      R^{[D]} : S \times A^\dag \to \PowerI(B^\dag \times S) \implies \sem{R^{[D]}}^{\PowerI} : S \to \PowerI\bigl((A^\dag)^\omega \causto (B^\dag)^\omega\bigr)
    \end{equation*}
  \end{NEWENV}

  The question can of course be hard to decide in general.  We give
  no algorithm or guarantees for arbitrary, undisciplined use of
  partiality and nondeterminism.  However, if these features are used
  in the sense of total functional programming, with partiality within
  each pattern matching rule, and harmless static nondeterminism
  between non-overlapping alternatives, then the problem reduces to
  standard analyses of pattern matching.
\end{enumerate}
\begin{NEWENV2}
  See Figure~\ref{fig:sah-rel} below for a worked-out example.
\end{NEWENV2}

\begin{NEWENV}
  To abstract from initial states is a typical tactic of the
  coalgebraic approach, in contrast to classical automata theory.  In
  a concrete program, of course, an initial value needs to be supplied
  or inferred.  We envisage both delay nodes annotated with their
  initial state, and thus equivalent to the traditional
  \emph{followed-by} data-flow operator, and default values such as
  zero for numeric types.
\end{NEWENV}

\subsection{Laws of Composition}
\label{composition}

\begin{NEWENV}
The three axes of relational composition should correspond to basic
operations on the stream function level: \footnote{
\begin{NEWENV3}%
  See \cite{Spivey1988} for the Z notation for set comprehensions used here and below.
\end{NEWENV3}}
\begin{align*}
  \sem{R \mathbin\| S}^M(s, t) &= \{ f : \sem{S}^M(s); g :
  \sem{S}^M(t) \mathrel\bullet f \otimes g \}
  \\
  \sem{R \cdot S}^M(s, t) &= \{ f : \sem{R}^M(s); g : \sem{S}^M(t)
  \mathrel\bullet g \circ f \}
  \\
  \sem{\underbrace{R \fatsemi \dots \fatsemi R}_{n>0}}^M(s) &= \{ f :
  \sem{R}^M(s) \mathrel\bullet f^{(n)} \}
\end{align*}
While these laws are readily proven for $M = \PowerI$ by equivalence
to ordinary causal stream functions and coinduction, the situation for
$M = \Power, \Power*$ is more complicated.  We leave these as
conjectures for future validation, or falsification, of our
theoretical framework.  The simplest case is the most important for
program structure in the large, anyway, as discussed in
section~\ref{compositional} above.

Note that the last rule, although less general than the preceding two,
is of great practical importance, because it enables the code
optimization technique known as
\begin{NEWENV4}
  \emph{strip mining} (block-wise \emph{loop unrolling}),
\end{NEWENV4}
which is essential for buffer-based high-performance implementations
of stream functions.
\end{NEWENV}

\section{Semantic Strategy}
\label{strategy}

As we have just discussed, element-level relational semantics are the
stepping stone for stream function semantics.  Thus we need to assign
stateful relations to programs, preferrably by induction over the
expression structure.  To this end, we shall give a core program
calculus and suitable normalization algorithms.  They reduce program
expressions in either functional or data-flow style to a logical form,
which essentially names interface variables and lists primitive
computation operations as assignments to their respective target
variables.  The intended meaning is then the largest relation on the
possible values of interface variables that satisfies the assignments
as constraints.

Before the full formal details of inductive interpretation are given
in the next section, we briefly introduce the constraint building
blocks associated with primitive operations.  References to other
user-defined components are non-recursive. Hence they can be unfolded
by inlining at the semantic level, at do not require special
consideration.

\subsection{Data Flow}

The interpretation of data flow primitives is depicted in
Figure~\ref{fig:prim-data}.  Delay operations ($\delta$) are resolved as
discussed above.  Primitive operations on data values are stateless
functions ($f$), and lifted to the stateful relational level in the
obvious way.  Data term constructors are special among these
operations, as they are injective and have pairwise disjoint ranges.
They have partial inverse functions ($f^{-1}$) with pairwise disjoint
domains.  In order to handle partiality, we lift inverse constructors
to the relational level by adding a control output to indicate
definedness.  This is in particular necessary for nullary
constructors.

\begin{figure}
  \begin{gather*}
    \inference{f(x_1, \dots, x_n) = y_1, \dots, y_m}{{} \mathbin/ x_1,
      \dots, x_n \stackrel{f}\longmapsto y_1, \dots, y_m \mathbin/ {}}
    \qquad%
    \inference{f(x_1, \dots, x_n) = y}{{} \mathbin/ y
      \stackrel{f^{-1}}\longmapsto x_1, \dots, x_n, \top \mathbin/ {}}
    \qquad%
    \inference{y \not\in \operatorname{ran} f}{{} \mathbin/ y
      \stackrel{f^{-1}}\longmapsto \underbrace{\bot, \dots, \bot}_n,
      \bot \mathbin/ {}}
    \\[-\medskipamount]
    \inference{}{s \mathbin/ x \stackrel{\delta}\longmapsto s \mathbin/ x}
  \end{gather*}
  \caption{Semantic building blocks (data flow)}
  \label{fig:prim-data}
\end{figure}

For programs written in a declarative front-end formalism, variables
of control type are not specified by the user, but arise only in this
synthetic way.  Whereas inverse constructors act as their sources,
special control flow primitives are introduced to act as their sinks.

\subsection{Control Flow}
\label{control-flow}

The interpretation of control flow primitives is depicted in
Figure~\ref{fig:prim-control}.  As we have already criticized in the
introduction, there is no canonical control flow construct in
data-flow graph approaches to programming.  Therefore we only consider
pattern matching,
\begin{NEWENV2}
  in the form of a \emph{case} expression, confer~\cite{Augustsson1985},
\end{NEWENV2}
as a universal control flow construct of
non-recursive functional programming.  To this end, we adapt the basic
idea of the SSA form for imperative programming to a more data-centric
scenario.

Results of alternative branches are joined using virtual ``Phony''
($\phi$) nodes, which are usually thought of as choosing the
appropriate case nondeterministically in the static perspective, but
deterministically in the dynamic perspective, because only one option
is available at each instance.  Unlike SSA, however, we do not assume
alternative branches to be taken by single-threaded execution of
conditional jumps, but instead we
\begin{NEWENV}
  exploit the absence of side effects in declarative programs, and
\end{NEWENV}
allow for speculative evaluation in parallel.  This gives more freedom
for implementation strategies, and may even make sense literally for
hardware or near-hardware back-ends.
\begin{NEWENV}
  In this view, $\phi$ nodes choose nondeterministically from their
  defined input, if any.
\end{NEWENV}

Choices from alternative values are then to be made from concurrent
pattern matching rules, with the definedness of each rule conditional
on the successful matching of \emph{all} its pattern constructors.
The control aspect of choice is reflected back to the data flow view
by our novel extension to SSA: ``Guard'' ($\gamma$) nodes propagate a
single data input, subject to strictness in one or more additional
control inputs.
\begin{NEWENV}
  They act as control flow sinks, corresponding to inverse
  constructors as sources, and perform the selection of (results~of)
  alternative branches.
\end{NEWENV}

\begin{NEWENV4}
  Note that $\gamma$ and $\phi$ nodes, applied to control inputs
  exclusively, reduce to Boolean conjunction and disjunction,
  respectively.
\end{NEWENV4}

\begin{figure}
  \begin{gather*}
    \begin{NEWENV}
      \inference{y \in \{x_1, \dots, x_n\} \setminus \{ \bot \}}{{}
        \mathbin/ x_1, \dots, x_n \stackrel{\phi}{\longmapsto} y
        \mathbin/ {}} \qquad \inference{\{x_1, \dots, x_n\} \subseteq
        \{ \bot \}}{{} \mathbin/ x_1, \dots, x_n
        \stackrel{\phi}{\longmapsto} \bot \mathbin/ {}}
    \end{NEWENV}
    \\[\medskipamount]
    \inference{\bot \not\in \{ c_1, \dots, c_k \}}{{} \mathbin/ x,
      c_1, \dots, c_k \stackrel{\gamma}{\longmapsto} x \mathbin/}
    \qquad \inference{\bot \in \{ c_1, \dots, c_k \}}{{} \mathbin/ x,
      c_1, \dots, c_k \stackrel{\gamma}{\longmapsto} \bot \mathbin/}
  \end{gather*}
  \caption{Semantic building blocks (control flow)}
  \label{fig:prim-control}
\end{figure}

\section{Core Calculus}
\label{core}

\begin{figure}
  \begin{align*}
    \boxed{
    \begin{aligned}
      \mathit{Expr} &::= () \bigm| \mathit{Expr} \mathbin,
      \mathit{Expr} \bigm| \mathit{Var} \bigm| \mathit{Val} \bigm|
      \mathit{Op}\; \mathit{Expr}
      \\
      &\quad\bigm| {} \mathrel{\mathbf{let}} \mathit{Vars} :=
      \mathit{Expr} \mathrel{\mathbf{in}} \mathit{Expr} \bigm| {}
      \mathrel{\mathbf{case}} \mathit{Expr} \mathrel{\mathbf{of}}
      \mathit{Rule}
      \\[\smallskipamount]
      \mathit{Vars} &::= () \bigm| \mathit{Vars} \mathbin,
      \mathit{Vars} \bigm| \mathit{Var}
      \\[\smallskipamount]
      \mathit{Op} &::= \mathit{Fun} \bigm| \mathit{Cons} \bigm|
      \mathit{Cons}^{-1} \bigm| \delta \bigm| \gamma \bigm| \phi
      \\[\smallskipamount]
      \mathit{Rule} &::= \mathit{Pat} \to \mathit{Expr} \bigm|
      \mathit{Rule} \sqcup \mathit{Rule}
      \\[\smallskipamount]
      \mathit{Pat} &::= () \bigm| \mathit{Pat} \mathbin, \mathit{Pat}
      \bigm| \mathit{Var} \bigm| \mathit{Cons} \; \mathit{Pat}
      \\[\smallskipamount]
      \mathit{Abs} &::= \operatorname\lambda \mathit{Rule} \bigm| [
      \mathit{Face} \mathrel{\mathbf{where}} \mathit{Form} ]
      \\[\smallskipamount]
      \mathit{Face} &::=
      \quadri{\mathit{Vars}}{\mathit{Vars}}{\mathit{Vars}}{\mathit{Vars}}
      \\[\smallskipamount]
      \mathit{Form} &::= \top \bigm| \bot \bigm| \mathit{Form} \land
      \mathit{Form} \bigm| \mathit{Form} \lor \mathit{Form}
      \\
      &\quad\bigm| \operatorname\exists \mathit{Var} \mathpunct{}
      \mathit{Form} \bigm| \mathit{Vars} := \mathit{Expr} \bigm|
      \mathit{Var} = \bot \bigm| \mathit{Var} \neq \bot
    \end{aligned}}
    \qquad
    \begin{array}{llc@{\enspace}c@{\enspace}c}
      \toprule
      && \mathbf 1 & \mathbf 2 & \mathbf 3
      \\ \midrule
      \multirow{4}{*}{$\mathit{Expr}$} & \mathit{Expr} \mathbin, \mathit{Expr}  & \yes & \some & \some
      \\
      & \mathit{Op}\;\mathit{Expr} & \yes & \some & \some
      \\
      & \mathrel{\mathbf{let}} \cdots \mathrel{\mathbf{in}} \cdots & \yes & \no & \no
      \\
      & \mathrel{\mathbf{case}} \cdots \mathrel{\mathbf{of}} \cdots & \yes & \no & \no
      \\ \midrule
      \multirow{3}{*}{$\mathit{Op}$} & \mathit{Cons}^{-1} & \no & \yes & \yes
      \\
      & \delta & \yes & \no & \no
      \\
      & \gamma \bigm| \phi & \no & \yes & \no
      \\ \midrule
      \mathit{Rule} & & \yes & \no & \no
      \\
      \mathit{Pat} & & \yes & \no & \no
      \\ \midrule
      \mathit{Abs} & \operatorname\lambda \mathit{Rule} & \yes & \no & \no
      \\ \midrule
      \multirow{2}{*}{$\mathit{Form}$} & \bot \bigm| \mathit{Form} \lor \mathit{Form} & \no & \no & \yes
      \\
      & \mathit{Var} = \bot \bigm| \mathit{Var} \neq \bot & \no & \no & \yes
      \\ \bottomrule
    \end{array}
  \end{align*}
  \caption{Core calculus abstract syntax. {\itshape Left} -- Grammar
    rules.  {\itshape Right} -- Special forms (synopsis):
    allowed~$\yes$; restricted~$\some$; forbidden~$\no$.}
  \label{fig:absy}
\end{figure}

The abstract syntax of the core calculus is depicted in
Figure~\ref{fig:absy} (left).  Many of the nonterminals and
productions are self-explanatory.
\begin{itemize}
\item Issues of concrete syntax, such as operator precedence, are not
  considered.  Various brackets are added in concrete examples
  for disambiguation.
\item Sufficient and distinct stores of variables, defined function
  names, and constructor names are referred to by $\mathit{Var}$,
  $\mathit{Fun}$ and $\mathit{Cons}$, respectively.
\item Literals $\mathit{Val}$ denote some primitive values of
  interest, including the undefined value $\bot$ as discussed in
  sections~\ref{compositional} and \ref{strategy}.
\item The nonterminals $\mathit{Expr}$, $\mathit{Vars}$ and
  $\mathit{Pat}$ are meant to form free monoids with associative
  operator $(,)$ and neutral element $()$.
\item Operators of the form $\mathit{Cons}^{-1}$, $\gamma$, $\delta$
  and $\phi$ are special, in the sense that they do not belong to the
  ordinary functional programming fragment of the language; they are
  explained in detail below.
\item The rule combinator $\sqcup$ denotes general nondeterministic
  choice.  It is nevertheless useful in deterministic programs to
  combine mutually exclusive rules. It follows that no resolution rule
  for overlapping patterns, such as first-fit or best-fit, is implied.
  An asymmetric first-fit combinator can be added with little
  difficulty, but is omitted here for simplicity.
\item The square bracket form of function abstraction, reminiscent of
  data-flow boxes, is a symmetric variant of $\lambda$ that names
  outputs as well as inputs.  It is more suitable for the direct
  denotation of data-flow graphs, in particular where the flow is not
  tree-shaped or outputs are subject to delayed feedback.
\item The notation for the nonterminal $\mathit{Face}$, specifying
  interface variables, mimics the tuple notation for stateful
  relations.
\end{itemize}

\noindent The notation is subject to a few static sanity conditions on
the use of variables:
\begin{itemize}
\item \emph{Linearity} -- Variables left of the arrow of a
  $\mathit{Rule}$ or $\mathit{Face}$ are assumed to be pairwise
  distinct.
\item \emph{Barendregt convention} -- Variables occuring in a
  $\mathit{Pat}$ or bound by $\exists$ in a $\mathit{Form}$ are
  assumed to be pairwise distinct, and disjoint from the free
  variables.
\item \emph{Single assignment} -- Variables assigned to by $:=$ in an
  $\mathit{Expr}$ or $\mathit{Form}$ are assumed to be pairwise
  distinct.
\item \emph{Determinism} -- In a concrete program, all variables in an
  $\mathit{Abs}$, except for specified pre-states and inputs, are
  assigned to by $:=$.  Note that, in a loose specification, an
  unassigned variable is understood to take \emph{all} possible values
  instead; hence we can simply take the intersection of overlapping
  partial specifications.
\end{itemize}
Note that $\mathbf{let}$ expression may be recursive at face value.
That the corresponding, apparently circular data flow is properly
causal shall be ensured after a suitable normalization,
see~\ref{form2} below.

The comma monoid structure of expressions, variables and patterns
caters for the inherent parallel compositionality of data flow, both
in terms of wires grouped into ``buses'', and boxes with varying arity
or juxtaposed.  This static \emph{shape} information is considered
distinct from (product) \emph{type} information.  The former is
treated in the following semantic rule system
\begin{NEWENV}
  (as usual; confer~\cite{Broy1988}),
\end{NEWENV}
whereas the latter is ignored for simplicity:

Arities of all operations are assumed to be statically known, and
combine associatively; for instance, given the arity information $f :
3 \to 2$, $g : 2 \to 1$ and $h : 0 \to 2$, the assignment $y, z :=
f\bigl(g(w, x), h()\bigr)$ is well-shaped.  Drawing the corresponding
data-flow graph is left as an exercise to the reader.  Note that named
variables always have arity $1$.

By contrast, we do not assume a particular system of data types, but
instead provide for generic term-shaped data, by expressions of the
form $\mathit{Cons}(\mathit{Expr})$.  This data universe is general
enough to accommodate most first-order data structures, and to support
pattern matching as a universal control-flow construct.

The large-scale structure of programs beyond the expression level is
assumed to be extremely simple:
\begin{itemize}
\item We assume that no higher-order or nested functions are present.
  If they are supported by a high-level front-end notation, they
  should be eliminated by lambda lifting and defunctionalization
  first.  This simplifies the semantic discussion greatly, but is also
  supported by pragmatic considerations; for instance, it is far from
  clear what a stream function-valued stream function actually means.
\item We do not allow for circular references among user-defined
  functions.  This rules out all forms of explicit recursion.  This is
  not as much of a loss as one might assume, since only
  \emph{instantaneous} recursion within the computation of a single
  stream element is affected; a feature that stream algorithms rarely
  require, in particular in real-time environments.
\item Thus the basic form of a closed program is a sequence of
  defining pairs $\mathit{Fun} = \mathit{Abs}$, where right-hand-side
  $\mathit{Fun}$ references are only to preceding definitions.
\end{itemize}

We define three distinct special forms as inductive sublanguages of
the full core language,
\begin{NEWENV}
  depicted synoptically in Figure~\ref{fig:absy} (right),
\end{NEWENV}
as stages of a program transformation that
achieves the unification of the functional paradigm with the data-flow
paradigm on one hand, and the front-end paradigms with the relational
back-end paradigm on the other.
\begin{NEWENV}
  The three forms focus on functional front-end, SSA or flat data-flow
  graph, and pure first-order logic, respectively.
\end{NEWENV}
We give translation rules between the forms as syntax-directed
rewriting algorithms.

As a running example of the forms and translations, we shall use the
\emph{sample-and-hold} component.  This small stream processing
component has a single signal input $x$ and output $y$, and either
feeds the current input through (\emph{sample}), or retains the
previous output (\emph{hold}), depending on some switching condition,
typically a trigger input $t$ taking values from
\begin{NEWENV4}
  the enumeration type
\end{NEWENV4}
$\{ \mathsf{S}, \mathsf{H} \}$.  The sample-and-hold component plays
an important role in many low-level stream computations, in particular
for audio signals.  It has the charming property that each of the
features that we propose to add to traditional data-flow programming,
is used in a minimal but nontrivial way.
\begin{NEWENV2}
  The only downside is that the example makes only trivial use of
  pattern matching, but readers can surely extrapolate from their
  experience with more complex case distinctions.
\end{NEWENV2}

To illustrate hierarchical specification,
implicit state and compositionality, Figure~\ref{fig:sah} depicts a
variety of views on the component: a loose nondeterministic
specification that abstracts from the switching condition, with either
implicit or explicit state, as well as a refined deterministic
component fragment that presupposes switching, and the corresponding
pattern matching front-end.  The concrete program arises by
composition of the latter two.  It is given in each of the three
special forms of our core calculus in turn, in a form that arises by
the given canonical translations up to straightforward
algebraic--logic simplifications.
\begin{NEWENV4}
  Note that the initial, loose specification is not in either of the
  special forms, but uses an ad-hoc, convenient mixture (with implicit
  state and $\phi$ nodes); this expressive and natural example
  justifies our treatment of the various transformation stages as
  subsets of a common language.
\end{NEWENV4}

\begin{figure}
  \centering
  \begin{pgfpicture}{0cm}{-0.25cm}{4cm}{2.75cm}
    \begin{sigbound}{\pgfxy(0.75,-0.25)}{\pgfxy(2.5,3)}
      \sigbox{\pgfxy(1.5,0)}{\pgfxy(1,1)}{\phi}
      \sigbox{\pgfxy(1.5,1.5)}{\pgfxy(1,1)}{\delta}
      \sigline{\pgfxy(0.5,0.25)}{\pgfxy(1.5,0.25)}
      \sigline{\pgfxy(2.5,0.5)}{\pgfxy(3.5,0.5)}
      \sigline{\pgfxy(1.5,2)}[\pgfxy(1,2)][\pgfxy(1,0.75)]{\pgfxy(1.5,0.75)}
      \sigline*{\pgfxy(3,0.5)}[\pgfxy(3,2)]{\pgfxy(2.5,2)}
    \end{sigbound}
    \sigport{w}{\pgfxy(0.5,0.25)}{x}
    \sigport{e}{\pgfxy(3.5,0.5)}{y}
  \end{pgfpicture}
  \hspace{1cm}
  \raisebox{-0.75cm}{
  \begin{pgfpicture}{0cm}{-1cm}{5cm}{3.5cm}
    \begin{sigbound}{\pgfxy(0.75,-0.25)}{\pgfxy(3.5,3)}
      \sigbox{\pgfxy(1.5,0)}{\pgfxy(1,1)}{\phi}
      \sigboxdiag*{\pgfxy(2.5,1.5)}{\pgfxy(1,1)}
      \sigline{\pgfxy(0.5,0.25)}{\pgfxy(1.5,0.25)}
      \sigline{\pgfxy(2.5,0.5)}{\pgfxy(4.5,0.5)}
      \sigline{\pgfxy(2.5,2)}[\pgfxy(1,2)][\pgfxy(1,0.75)]{\pgfxy(1.5,0.75)}
      \sigline*{\pgfxy(4,0.5)}[\pgfxy(4,2)]{\pgfxy(3.5,2)}
      \sigline{\pgfxy(3,3)}{\pgfxy(3,2.5)}
      \sigline{\pgfxy(3,1.5)}{\pgfxy(3,-0.5)}
    \end{sigbound}
    \sigport{w}{\pgfxy(0.5,0.25)}{x}
    \sigport{e}{\pgfxy(4.5,0.5)}{y}
    \sigport{n}{\pgfxy(3,3)}{s}
    \sigport{s}{\pgfxy(3,-0.5)}{s'}
  \end{pgfpicture}}
  \hspace{1cm}
  \raisebox{-0.75cm}{
  \begin{pgfpicture}{0cm}{-1cm}{4cm}{2cm}
    \begin{sigbound}{\pgfxy(0.75,-0.25)}{\pgfxy(2.5,1.5)}
      \sigbox{\pgfxy(1.5,0)}{\pgfxy(1,1)}{\phi}
      \sigline{\pgfxy(0.5,0.25)}{\pgfxy(1.5,0.25)}
      \sigline{\pgfxy(2.5,0.5)}{\pgfxy(3.5,0.5)}
      \sigline{\pgfxy(1,1.5)}[\pgfxy(1,0.75)]{\pgfxy(1.5,0.75)}
      \sigline*{\pgfxy(3,0.5)}{\pgfxy(3,-0.5)}
    \end{sigbound}
    \sigport{w}{\pgfxy(0.5,0.25)}{x}
    \sigport{e}{\pgfxy(3.5,0.5)}{y}
    \sigport{n}{\pgfxy(1,1.5)}{s}
    \sigport{s}{\pgfxy(3,-0.5)}{s'}
  \end{pgfpicture}}
  \\[5mm]
    \begin{pgfpicture}{-2cm}{-0.75cm}{2.5cm}{2.5cm}
    \begin{sigbound}{\pgfxy(-1.25,-0.75)}{\pgfxy(2.5,3.25)}
      \sigbox{\pgfxy(-0.5,-0.25)}{\pgfxy(1,1)}{\mathsf{S}^{-1}}
      \sigbox{\pgfxy(-0.5,1.25)}{\pgfxy(1,1)}{\mathsf{H}^{-1}}
      \sigline{\pgfxy(-1.5,-0.5)}{\pgfxy(1.5,-0.5)}
      \sigline{\pgfxy(-1.5,1)}[\pgfxy(-1,1)][\pgfxy(-1,1.75)]{\pgfxy(-0.5,1.75)}
      \sigline*{\pgfxy(-1,1)}[\pgfxy(-1,0.25)]{\pgfxy(-0.5,0.25)}
      \begin{sigctrl}
        \sigline{\pgfxy(0.5,0.25)}{\pgfxy(1.5,0.25)}
        \sigline{\pgfxy(0.5,1.75)}{\pgfxy(1.5,1.75)}
      \end{sigctrl}
    \end{sigbound}
    \sigport{w}{\pgfxy(-1.5,-0.5)}{x}
    \sigport{w}{\pgfxy(-1.5,1)}{t}
    \sigport{e}{\pgfxy(1.5,-0.5)}{x}
    \sigport{e}{\pgfxy(1.5,0.25)}{c}
    \sigport{e}{\pgfxy(1.5,1.75)}{d}
  \end{pgfpicture}
  \hspace{1cm}
  \begin{pgfpicture}{0cm}{-1cm}{6cm}{2.75cm}
    \begin{sigbound}{\pgfxy(0.75,-1)}{\pgfxy(4.5,3.75)}
      \sigbox{\pgfxy(3.5,0)}{\pgfxy(1,1)}{\phi}
      \sigbox{\pgfxy(3.5,1.5)}{\pgfxy(1,1)}{\delta}
      \sigbox{\pgfxy(1.5,-0.75)}{\pgfxy(1,1)}{\gamma}
      \sigbox{\pgfxy(1.5,0.75)}{\pgfxy(1,1)}{\gamma}
      \sigline{\pgfxy(0.5,-0.5)}{\pgfxy(1.5,-0.5)}
      \sigline{\pgfxy(2.5,-0.25)}[\pgfxy(3,-0.25)][\pgfxy(3,0.25)]{\pgfxy(3.5,0.25)}
      \sigline{\pgfxy(2.5,1.25)}[\pgfxy(3,1.25)][\pgfxy(3,0.75)]{\pgfxy(3.5,0.75)}
      \sigline{\pgfxy(4.5,0.5)}{\pgfxy(5.5,0.5)}
      \sigline{\pgfxy(3.5,2)}[\pgfxy(1,2)][\pgfxy(1,1.5)]{\pgfxy(1.5,1.5)}
      \sigline*{\pgfxy(5,0.5)}[\pgfxy(5,2)]{\pgfxy(4.5,2)}
      \begin{sigctrl}
        \sigline{\pgfxy(0.5,0)}{\pgfxy(1.5,0)}
        \sigline{\pgfxy(0.5,1)}{\pgfxy(1.5,1)}
      \end{sigctrl}
    \end{sigbound}
    \sigport{w}{\pgfxy(0.5,0)}{c}
    \sigport{w}{\pgfxy(0.5,-0.5)}{x}
    \sigport{w}{\pgfxy(0.5,1)}{d}
    \sigport{e}{\pgfxy(5.5,0.5)}{y}
  \end{pgfpicture}

  \begin{align*}
    \mathit{sah\_prog}_1 &= \bigl[x, t \to y \mathrel{\mathbf{where}}
    y := {}\mathrel{\mathbf{case}} t \mathrel{\mathbf{of}} \{
    \mathsf{S}() \to x ~\sqcup~ \mathsf{H}() \to \delta(y) \} \bigr]
    \\
    \mathit{sah\_prog}_2 &= \left[s \mathbin/ x, t \to y \mathbin/ y \mathrel{\mathbf{where}}
      \operatorname\exists c, d, v, w
      \left(
        \begin{aligned}
          c &:= \mathsf{S}^{-1}(t) & v &:= \gamma(x, c)
          \\
          d &:= \mathsf{H}^{-1}(t) & w &:= \gamma(s, d)
        \end{aligned}
        \quad y := \phi(v, w)
      \right)
    \right]
    \\
    \mathit{sah\_prog}_3 &= \left[s \mathbin/ x, t \to y \mathbin/ y \mathrel{\mathbf{where}}
      \operatorname\exists c, d, v, w
      \left(
        \begin{aligned}
          c &:= \mathsf{S}^{-1}(t) & (c \neq \bot \land v :=
          x) &\lor (c = \bot \land v := \bot)
          \\
          d &:= \mathsf{H}^{-1}(t) & (d \neq \bot \land w :=
          s) &\lor (d = \bot \land w := \bot)
          \\
          && (y := v &\lor y := w)
          \\
          && (v = \bot \lor w &= \bot \lor y \neq \bot)
        \end{aligned}
         \right)
    \right]
  \end{align*}
  \caption{Sample and hold.  {\em Top} -- loose specification with
    implicit ({\em left}) and explicit state and eliminated delay,
    before ({\em center}) and after ({\em right}) copy propagation.
    {\em Center} -- functionally partial implementation with external
    control ({\em right}) and frontend for functionally complete
    implementation with internal control ({\em left}).  {\em Bottom}
    -- Concrete program in first, second and third form,
    respectively.}
  \label{fig:sah}
\end{figure}

\newpage
\subsection{First Form}

A program is in \emph{first} form, if and only if the following
constraints are satisfied:
\begin{itemize}
\item No operation symbol of the form $\mathit{Cons}^{-1}$, $\gamma$
  or $\phi$ occurs in expressions.
\item Interface declarations $\mathit{Face}$ are of the form
  $\mathit{Vars}$, and do not mention state variables.
\item Logical formulas use the operators $\top$, $\land$, $:=$ and
  $\exists$ only.
\end{itemize}

This is the form of elementary functional and data-flow programming.
It is already more low-level than one would like for a real
programming or specification language front-end, although many more
advanced features can be reduced to this form by standard program
transformations, as discussed above.  It already allows to mix the
functional and data-flow paradigms compositionally.  

A corresponding instance of the running example is depicted in
Figure~\ref{fig:sah}.  It uses the data-flow box abstraction instead
of functional $\lambda$, because of the very common pattern that uses
delayed outputs as inputs in a feedback loop, which is slightly
awkward to express in functional style due to anonymous results.
Internally the behavior of the box is defined in terms of functional
pattern matching, which makes for a much more natural expression than
any ad-hoc data-flow switching mechanism,
\begin{NEWENV}
  and scales better to nontrivial case distinctions than if-then-else.
\end{NEWENV}
State is still implicit,
the appearance of element-wise computation is maintained.

\subsection{Second Form}
\label{form2}

A program is in \emph{second} form, if and only if the following constraints
are satisfied:
\begin{itemize}
\item No $\mathbf{let}$ or $\mathbf{case}$ construct appears in
  expressions.
\item No operation symbol of the form $\delta$ occurs in expressions.
\item Interface declarations $\mathit{Face}$ are of the form
  $\mathit{Vars} \mathbin/ \mathit{Vars}$, and do mention state
  variables.
\item Logical formulas use the operators $\top$, $\land$, $:=$ and
  $\exists$ only.
\item Expressions are flat: any term of the form
  $\mathit{Op}(\mathit{Expr})$ is actually of the simpler form
  $\mathit{Op}(\mathit{Var}^{*})$.
\item Assignments are atomic: the expression operator $(,)$ does not
  occur in right operands of $:=$.
\item Assignments are non-circular.
\end{itemize}

This is the basic exchange format for program analysis and code
generation.  State is made explicit, putting all computationally
relevant information together.  Structure is reduced to atomic
assignments, either copying values or storing results of a single
operation.  The form is inspired by standard form, such as A-normal
form of functional programs, or SSA form of imperative programs.  It
adopts the $\phi$~nodes for converging data flow from the latter, but
uses a different encoding of control flow, to be explained below.

\begin{NEWENV}
  A corresponding instance of the running example is depicted in
  Figure~\ref{fig:sah}.  It has two inverse constructors as sources of
  control flow, two $\gamma$ nodes as the corresponding sinks and one
  $\phi$ node for choice.  It also corresponds directly to the
  horizontal composition of the two fragments depicted in the center
  row of Figure~\ref{fig:sah}.
\end{NEWENV}

\begin{NEWENV}
  From the second form, data-flow graphs can be drawn, by reading the
  conjunction of assignments as a specification of adjacency, and
  element-level relational semantics can be read off by taking
  formulas as constraints in standard first-order logic.
\end{NEWENV}

\begin{NEWENV2}
  For the running example, relational element-level semantics and
  their coinductive lifting to streams are depicted in
  Figure~\ref{fig:sah-rel}.  The definition of relation $R$
  corresponds directly to the second form of the sample-and-hold
  program.  It has not been simplified in any way, in order to make
  the structure stand out.  The subsequent relation $R^\dag$ shows the
  removal of bottom values.  Primitive relations have been eliminated,
  by substituting their definitions, or alternatively by translation
  to third form; see below.

  Additionally, stream function semantics are given, deterministically
  for the concrete program, and nondeterministically for the loose
  specification depicted in the top row of Figure~\ref{fig:sah},
  respectively.  For the former, the state space is restricted, in the
  sense discussed in Section~\ref{compositional}, by exclusion of
  $\bot$.  Note that, from the strongly formal viewpoint, these
  self-referential presentations are a step backwards; to verify that
  thay are well-defined implicitly invokes coinduction.
\end{NEWENV2}

\begin{figure}
  \begin{NEWENV}
  \begin{align*}
    R &= \begin{aligned}[t] \bigl\{ &s, v, w, x, y : A \uplus \{ \bot
      \}; t : \{ \mathsf{S}, \mathsf{H}, \bot \} ; c, d : \{ \top,
      \bot \}
      \\
      &\bigm| {} \mathbin/ t \stackrel{\mathsf{S}^{-1}}\mapsto c
      \mathbin/ {} \land {} \mathbin/ t
      \stackrel{\mathsf{H}^{-1}}\mapsto d \mathbin/ {} \land {}
      \mathbin/ x, c \stackrel{\gamma}\mapsto v \mathbin/ {} \land {}
      \mathbin/ s, d \stackrel{\gamma}\mapsto w \mathbin/ {} \land {}
      \mathbin/ v, w \stackrel{\phi}\mapsto y \mathbin/ {}
      \\
      &\mathrel\bullet s \mathbin/ x, t \mapsto y \mathbin/ y \bigr\}
    \end{aligned}
    \\
    R^\dag &= \bigl\{ s, x, y : A; t : \{ \mathsf{S},
    \mathsf{H} \} \bigm| (t = \mathsf{S} \Rightarrow y = x) \land (t =
    \mathsf{H} \Rightarrow y = s) \mathrel\bullet s \mathbin/ x, t \mapsto y
    \mathbin/ y \bigr\}
  \end{align*}
  \begin{align*}
    \sem{R^{[A]}}^{\PowerI}(s) &= \{ \begin{NEWENV4}h(s)\end{NEWENV4}
    \} \quad\text{with}\quad \begin{NEWENV4} h(s)\bigl(\mathit{cons}((x, t), a)\bigr)
      = \mathit{cons}\bigl(y, h(y)(a)\bigr)\end{NEWENV4}
    &&\text{where} & y &=
    \begin{cases}
      x & \text{if~} t = \mathsf{S}
      \\
      s & \text{if~} t = \mathsf{H}
    \end{cases}
    \\
    \sem{Q^\dag}^{\Power}(s) &= \Bigl\{ h : A^\omega \causto A^\omega
    \Bigm| \bigl(\operatorname\forall a : A^\omega; n : \mathbb{N}
    \mathrel\bullet h(a)_n \in \{ a_n, a'_n \}
    \bigr) \Bigr\}
    &&\text{where} & a' &= \mathit{cons}\bigl(s, h(a)\bigr)
  \end{align*}
  \caption{Relation semantics of sample-and-hold.  {\itshape Top} --
    deterministic element-level; first partial (compositional), then
    total (boundary), simplified.  {\itshape Bottom} -- stream-level;
    first deterministic program, then nondeterministic specification.}
  \label{fig:sah-rel}
\end{NEWENV}
\end{figure}

\subsection{Reduction from First to Second Form}

A simple production-wise comparison of the first and second form
reveals the tasks to be solved by a translation:
\begin{enumerate}
\item State must be made explicit by elimination of $\delta$
  operations and introduction of state variables.  The solution idea
  is already depicted in Figure~\ref{fig:rel}.  The data flow of the
  component must become non-circular in the process, otherwise the
  source program is ill-formed.
\item Composite applicative expressions must be decomposed into their
  operations, by explicitly binding all intermediate values to
  variables.
\item Variable-binding expressions and abstractions, and hence pattern
  matching rules, must be eliminated entirely.  This entails a
  translation of patterns to applications by a reversal of data flow,
  replacing $\mathit{Cons}$ with $\mathit{Cons}^{-1}$.
\end{enumerate}

Both first and second form share a simple structure of logical
formulas.  Since existential quantification can always be lifted from
a conjunction, thanks to the Barendregt convention even without
renaming, they admit a prenex normal form: A (commutative) sequence of
existential quantifiers introducing local variables, followed by a
(also commutative, thanks to single assignment) sequence of
assignments.

The reduction algorithm specified below is given in terms of
syntax-directed rewrite rules.  It is slightly unconventional, in the
sense that context information is collected strictly bottom-up.  Hence
contexts appear on the right rather than the left hand sides of
rewrite rules, and specify variables contributed by the translation
target, rather than depended upon by the translation source.  A few
details of the notation need explanation:
\begin{itemize}
\item Meta-level variables $c, d$ and $s, \dots, z$ stand for vectors
  of zero or more pairwise distinct object-level variables,
  \begin{NEWENV4}
    the former of control and the latter of data type, respectively.
  \end{NEWENV4}
  Individual variables are selected by subscripts if necessary.
  Following the tradition of linear algebra, subscripts $i,j$ range
  from $1$ to $m,n$, respectively.  Other meta-variables stand for
  singular language fragments.
\item Static checking of shape information boils down to comparisons
  of the length of variable vectors.  We write $x : n$ to state that
  vector $x$ contains $n$ variables, and $x \sim y$ to state that
  vectors $x$ and $y$ contain the same number of variables.
\item Judgements with bottom-up context, making up the main
  antecedents and the conclusions of rewrite rules, take the form $P
  \rewrite \Gamma \vdash Q$, where $P,Q$ are fragments of the
  source/target form, respectively, possibly belonging to different
  nonterminals, and $\Gamma$ is a collection of free variables of $Q$,
  sorted into different roles.  In order to distinguish the roles, we
  adapt
  the interface notation of stateful relations from section~\ref{rel}:
  $s \mathbin/ x \to y \mathbin/ s'
  $ has variables $s, x, y, s'
  $ for pre-state, input, output, and
  post-state,
  respectively.  In order to reduce typographical clutter, we omit
  parts of this sequence, separators and all, if they contain zero
  variables, according to the following nesting, indicated by
  horizontal lines:
  \begin{equation*}
    \overline{\underline{s
        \mathbin/ {}}~ x \to{}}~ y ~\underline{{}\mathbin/ s'} ~
  \end{equation*}
\end{itemize}

The naive bottom-up rewriting strategy has the advantage that the
problem of static shape inference is solved automatically.  The minor
downside is that, since no binding information is propagated
downwards, the resulting algorithm tends to produce redundant copies
of variables.  Similar effects are known from naive SSA
algorithms~\cite{Cytron1991}.  This is not an issue unless one
actually builds a compiler.  Results can of course be simplified with
a standard copy propagation pass.  To the contrary, the default
strategy to create fresh variables bottom-up has nice properties from
the perspective of algorithmic analysis, wich are easy to verify on a
rule-wise basis:
\begin{enumerate}
\item The sanity conditions on variable use are preserved.
\item Variables introduced in translation are either
  $\exists$-quantified locally, or propagated upwards in the context.
\end{enumerate}

\begin{figure}
  \begin{gather*}
    \inference{}{() \rewrite \ejudge{}{}{}{\top}}[Unit] \qquad
    \inference{s \sim s' & t \sim t' \\ a \rewrite
      \ejudge{s}{x}{s'}{A} & b \rewrite \ejudge{t}{y}{t'}{B}} {a, b
      \rewrite \ejudge{s,t}{x,y}{s',t'}{A \land B}}[Agg]
    \\[\medskipamount]
    \inference{x \sim y : 1 & y \text{~fresh}}{x \rewrite
      \ejudge{}{y}{}{x := y}}[Ref] \quad \inference{s \sim s' & x : n
      & y : m & t \sim t' : k & t, t', y \text{~fresh}
      \\
      f : k \mathbin/ n \to m \mathbin/ k & a \rewrite
      \ejudge{s}{x}{s'}{A}}{f(a) \rewrite
      \ejudge{s,t}{y}{s',t'}{\operatorname\exists x \bigl(A \land y
        \mathbin/ t' := f(t \mathbin/ x)\bigr)}}[App]
    \\[\medskipamount]
    \inference{s \sim s' & x \sim y \sim t \sim t' : m & t, t', y
      \text{~fresh}
      \\
      a \rewrite \ejudge{s}{x}{s'}{A} & B \equiv \bigwedge_i (y_i :=
      t_i \land t'_i := x_i)}{\delta(a) \rewrite
      \ejudge{s,t}{y}{s',t'}{\operatorname\exists x (A \land
        B)}}[Delay]
    \\[\medskipamount]
    \inference{s \sim s' & t \sim t' & x \sim y : n
      \\
      a \rewrite \ejudge{s}{x}{s'}{A} & b \rewrite
      \ejudge{t}{z}{t'}{B} & C \equiv \bigwedge_j (y_j :=
      x_j)}{\mathrel{\mathbf{let}} y := a \mathrel{\mathbf{in}} b
      \rewrite \ejudge{s,t}{z}{s',t'}{\operatorname\exists x, y (A
        \land B \land C)}}[Let]
    \\[\medskipamount]
    \inference{s \sim s' & t \sim t' & x \sim y : n
      \\
      a \rewrite \ejudge{s}{x}{s'}{A} & r \rewrite
      \rjudge{t}{y}{z}{t'}{B} & C \equiv \bigwedge_j (y_j :=
      x_j)}{\mathrel{\mathbf{case}} a \mathrel{\mathbf{of}} r \rewrite
      \ejudge{s,t}{z}{s',t'}{\operatorname\exists x,y (A \land B \land
        C)}}[Case]
  \end{gather*}
  \caption{Core calculus reduction from first to second form
    (Expressions)}
  \label{fig:red-expr}
\end{figure}

The rewrite rules concerning expressions are depicted in
Figure~\ref{fig:red-expr}.  Expressions are generally rewritten to
logical formulas, with a bottom-up context of pre-states, outputs and
post-states, but no inputs.
\begin{itemize}
\item Rules (Unit) and (Agg) collect contexts and logical clauses.
  Associativity is easy to see.
\item Rule (Ref) creates a fresh copy of a variable inherited from
  the source.
\item Rules (App) and (Delay) decompose nested terms by introducing
  variables $x$ for the intermediate values, and $t, t'$ for the state
  of the referenced function.  Many primitive functions, including
  constructors, are stateless with $k = 0$.  $m$-ary delay is reduced
  to $2m$ copying assignments as depicted in Figure~\ref{fig:rel}.
\item Rules (Let) and (Case) are essentially data-flow sequential
  compositions.
\end{itemize}

\begin{figure}
  \begin{gather*}
    \inference{}{() \rewrite \pjudge{}{}{\top}}[Unit'] \qquad
    \inference{s \sim s' & t \sim t' \\ p \rewrite \pjudge{x}{c}{A} &
      q \rewrite \pjudge{y}{d}{B}} {p, q \rewrite \pjudge{x,y}{c,d}{A
        \land B}}[Agg']
    \\[\medskipamount]
    \inference{x \sim y : 1 & y \text{~fresh}}{x \rewrite \pjudge{y}{}{y :=
        x}}[Ref'] \qquad \inference{x : m & y \sim d : 1 & f : m \to 1 &
      y, d \text{~fresh} \\ p \rewrite \pjudge{x}{c}{A}}{f(p) \rewrite
      \pjudge{y}{c,d}{\operatorname\exists x \bigl(A \land x,d :=
        f^{-1}(y)\bigr)}}[App']
  \end{gather*}
  \caption{Core calculus reduction from first to second form
    (Patterns)}
  \label{fig:red-pat}
\end{figure}

The rewrite rules concerning patterns are depicted in
Figure~\ref{fig:red-pat}.  Patterns are generally rewritten to logical
formulas, with a bottom-up context of inputs
\begin{NEWENV4}
  and control outputs
\end{NEWENV4}
but no states.
Rules mimic their expression counterparts closely, except for (App'),
which introduces a control output $d$ to indicate the success of
matching against constructor $f$, and exploits the statelessness of
$f$ for simplification.

\begin{figure}
  \begin{gather*}
    \begin{NEWENV4}
      \inference{s \sim s' & y \sim z : n & z \text{~fresh}
        \\
        p \rewrite \pjudge{x}{c}{A} & a \rewrite \ejudge{s}{y}{s'}{B}
        & C \equiv \bigwedge_j \bigl(z_j := \gamma(y_j, c)\bigr) }{p
        \to a \rewrite \rjudge{s}{x}{z}{s'}{\operatorname\exists c, y
          (A \land B \land C)}}[Match]
    \end{NEWENV4}
    \\[\medskipamount]
    \begin{NEWENV4}
      \inference{s \sim s' & t \sim t' & u \sim w \sim y : m & v \sim
        x \sim z : n & y, z \text{~fresh}
        \\
        q \rewrite \rjudge{s}{u}{v}{s'}{A} & r \rewrite
        \rjudge{t}{w}{x}{t'}{B}
        \\
        C \equiv \bigwedge_i (u_i := y_i \land w_i := y_i) & D \equiv
        \bigwedge_j \bigl(z_j := \phi(v_j, x_j)\bigr) }{ q \sqcup r
        \rewrite \rjudge{s,t}{y}{z}{s',t'}{\operatorname\exists u,
          v, w, x (A \land B \land C \land D)}}[Choice]
    \end{NEWENV4}
  \end{gather*}
  \caption{Core calculus reduction from first to second form
    (Rules)}
  \label{fig:red-rule}
\end{figure}

The rewrite rules concerning pattern matching rules are depicted in
Figure~\ref{fig:red-rule}.  Rules are generally rewritten to logical
formulas
\begin{NEWENV4}
  with full bottom-up context.
\end{NEWENV4}

\newpage
\begin{itemize}
\item Rule (Match) plugs together the complementary context of left
  and right hand side.
  \begin{NEWENV4}
    Each data output of the right hand side is guarded with \emph{all}
    control outputs of the left hand side (formula $C$); this is
    necessary since there need not be a direct data dependence.
  \end{NEWENV4}
  \begin{NEWENV}
  \item 
    \begin{NEWENV4}
      Rule (Choice) composes subrules speculatively in parallel, by
      distributing the common inputs (formula $C$), and collecting the
      alternative outputs pairwise (formula $D$).
    \end{NEWENV4}


  \end{NEWENV}
\end{itemize}

\begin{figure}
  \begin{gather*}
    \inference{s \sim s'
      \\
      r \rewrite \judge{s}{x}{y}{s'}{}{A}}{\operatorname\lambda r
      \rewrite {}[\quadri{s}{x}{y}{s'} \mathrel{\mathbf{where}}
      A]}[Lambda]
    \\[\medskipamount]
    \inference{ s \sim s' & w \sim y : n & x \sim z : m
      \\
      A \rewrite \judge{s}{w}{x}{s'}{}{A'} & B \equiv \bigwedge_j (w_j
      := y_j) & C \equiv \bigwedge_i (z_i := x_i)}{[\quadri{}{y}{z}{}
      \mathrel{\mathbf{where}} A] \rewrite{} [\quadri{s}{y}{z}{s'}
      \mathrel{\mathbf{where}} \operatorname\exists w,x (A' \land B
      \land C)]}[Box]
  \end{gather*}
  \caption{Core calculus reduction from first to second form
    (Abstractions)}
  \label{fig:rew-abs}
\end{figure}

\begin{figure}
  \begin{gather*}
    \inference{}{\top \rewrite \judge{}{}{}{}{}{\top}}[True] \qquad
    \inference{s \sim s' & t \sim t'
      \\
      A \rewrite \judge{s}{w}{x}{s'}{}{A'} & B \rewrite
      \judge{t}{y}{z}{t'}{}{B'}}{A \land B \rewrite
      \judge{s,t}{w,y}{x,z}{t,t'}{}{A' \land B'}}[And]
    \\[\medskipamount]
    \inference{s \sim s' & x \sim y : n \\ a \rewrite \ejudge{s}{y}{s'}{A}
      & B \equiv \bigwedge_j (x_j := y_j)}{x := a \rewrite
      \judge{s}{}{x}{s'}{}{\operatorname\exists y(A \land B)}}[Assign]
    \qquad
    \inference{s \sim s'
      \\
      A \rewrite \judge{s}{x}{y}{s'}{}{A'}}{\operatorname\exists w\,
      A \rewrite \judge{s}{x}{y}{s'}{}{A'}}[Exists]
  \end{gather*}
  \caption{Core calculus reduction from first to second form (Formulas)}
  \label{fig:rew-form}
\end{figure}

\begin{NEWENV}
  The rewrite rules concerning component abstractions and logic
  formulas are depicted in Figures~\ref{fig:rew-abs} and
  \ref{fig:rew-form}, respectively.  Abstractions, being independent
  top-level program constructs, are rewritten to abstractions with no
  context.  Formulas are rewritten to formulas with full bottom-up
  context.
  All rules are fairly straightforward.  Note that rule (Exists)
  exploits the Barendregt convention to skip checks for variable
  capture, since $w$ is disjoint from $s,s',x,y$.
\end{NEWENV}

\subsection{Third Form}

A program is in \emph{third} form, if and only if the following
constraints are satisfied:
\begin{itemize}
\item No $\mathbf{let}$ or $\mathbf{case}$ construct appears in
  expressions.
\item No operation symbol of the form $\delta, \gamma, \phi$ occurs in
  expressions.
\item Interface declarations $\mathit{Face}$ are of the form
  $\mathit{Vars} \mathbin/ \mathit{Vars}$, and do mention state
  variables.
\item Expressions are flat: any term of the form
  $\mathit{Op}\;\mathit{Expr}$ is actually of the simpler form
  $\mathit{Op}(\mathit{Var}^{*})$.
\item Assignments are atomic: the expression operator $(,)$ does not
  occur in right operands of $:=$.
\end{itemize}

This is the fundamental logical form of component behavior.  It
contains no operations except for primitive functions and their
inverses; all administrative nodes have been eliminated.  Hence, on
the theoretical level, it serves as a demonstration of the ``virtual''
nature of the synthetic nodes we have introduced.

On the practical level, the third form is obviously less suited to
analyses and transformations that aim at program \emph{execution},
because properties such as single assignment are obscured.  On the
other hand, it may be found useful for other kind of analysis.  For
instance, the problem of well-defined, that is, complete and
non-overlapping, pattern matching rules can be solved in this form
using standard SAT techniques.
\begin{NEWENV4}
  All formulas introduced by translation rules for either phase are
  easily seen to be already in, or leading directly to, conjunctive
  normal form.
\end{NEWENV4}

\subsection{Reduction from Second to Third Form}

A comparison of the second and third form yields a single translation
task, namely the elimination of $\gamma$ and $\phi$ nodes.  In return,
disjunction and tests for $\bot$ can be added to logical formulas.

The translation rules are depicted in Figure~\ref{fig:rew-control}.
This translation is algorithmically much simpler than the preceding
one, because it involves no rearrangement of variables, and can hence
be presented in context-free form.  It is a straightforward
realization of the specification in Figure~\ref{fig:prim-control}.
Additional, omitted rules for deep rewriting are trivial.
\begin{NEWENV4}
  The correctness of these rules with respect to the semantics of
  $\gamma$ and $\phi$ nodes, given in Figure~\ref{fig:prim-control},
  is easy to verify.
\end{NEWENV4}


\begin{figure}
  \begin{gather*}
    \inference{x, y : 1 & c : m
      \\
      C \equiv (\bigwedge_{i=1}^m c_i \neq \bot) & D \equiv
      (\bigvee_i c_i = \bot)}{y := \gamma(x, c)
      \rewrite \begin{NEWENV4}(C \lor y := \bot) \land (D \lor y :=
        x)\end{NEWENV4}}[Guard]
    \\[\medskipamount]
    \inference{x : n & y : 1
      \\
      C \equiv (\bigvee_{\!j=1}^n y := x_j) & D \equiv
      \begin{NEWENV4}(\bigwedge_j x_j = \bot)\end{NEWENV4}}{y
      := \phi(x) \rewrite \begin{NEWENV4}C \land (D \lor y \neq
        \bot)\end{NEWENV4}}[Phony]
  \end{gather*}
  \caption{Core calculus reduction from second to third form}
  \label{fig:rew-control}
\end{figure}

\section{Conclusion}

\begin{NEWENV2}
  Starting from domain-specific and rigidly mathematical requirements,
  we have outlined an inter-paradigm design that unifies data-flow and
  total functional styles for stream programming.  We have given a
  formal semantical framework in terms of established concepts, namely
  (monadic) coinduction over set-theoretic coalgebras, and Z-style
  stateful step relations.  These concepts have been combined in a
  novel way, and mapped to a full-fledged core programming calculus,
  as the fundamental step of programming language design.

  The calculus has the expressivity to accomodate pure programs in
  either paradigm, as well as pragmatic and fine-grained mixtures.
  Transformation rules have been given to reduce programs to a
  constraint form where their relational semantics can be read off
  directly.  
  \begin{NEWENV3}
    The calculus is disciplined enough to scale from deterministic
    programs to nondeterministic descriptions, yet avoid the typical
    anomalies.
  \end{NEWENV3}
  We have demonstrated how to embed the data-flow view
  straightforwardly into a state-of-the-art declarative intermediate
  representation of programming languages (SSA).  Thereby we inherit a
  large body of knowledge concerning both reduction of front-end
  language features, and compilation to back-end platforms.
\end{NEWENV2}

\begin{NEWENV2}
  A core calculus with precise semantics may be a major and essential
  step towards the realization of a paradigm, but of course there
  remains a lot of future work.  

  On the theoretical level, the finer implications of set-theoretic
  monadic coinduction are still poorly understood.  We point to the
  unproven conjectures regarding the three axes of relational
  composition in Section~\ref{composition} as an open example problem.

  On the level of algorithmic expressivity, our strict prohibition of
  general recursion need to be reconsidered carefully.  Experience
  shows that, for most applications in the domains of interest,
  non-recursive algorithms with single-step delay nodes suffice: the
  coinductive lifting from element-wise to stream calculation does all
  the rest.  But in some illuminating corner cases, more powerful
  means of feedback may be required.

  On the level of programming language engineering, our approach is
  readily compatible with many standard procedures, both regarding
  front-end issues such as type systems and diagnostics, and back-end
  code generation techniques.  On the other hand, due to the rigidity
  of semantics, it steers clear of some notorious implementation
  difficulties, such as the non-compositionality of operation
  scheduling in the presence of arbitrary instantaneous feedback;
  confer~\cite{Caspi1987}.  That these hopes are indeed justified
  shall be demonstrated by a forthcoming compiler.  Another
  challenging open problem is the design of a type system that
  enforces our strong assumptions in a way that is flexible,
  unobtrusive and practically useful.
\end{NEWENV2}

\section*{Acknowledgments}

Signal graphs for the ARMA and ADSR models in Figures~\ref{fig:arma}
and \ref{fig:adsr}, respectively, have been drawn using the free
statistical software system~\textsf{R}.

\raggedright
\bibliographystyle{eptcs}
\catcode`\_=\active
\newcommand_{\_}
\bibliography{boxes}

\begin{thebibliography}{10}
\providecommand{\bibitemdeclare}[2]{}
\providecommand{\surnamestart}{}
\providecommand{\surnameend}{}
\providecommand{\urlprefix}{Available at }
\providecommand{\url}[1]{\texttt{#1}}
\providecommand{\href}[2]{\texttt{#2}}
\providecommand{\urlalt}[2]{\href{#1}{#2}}
\providecommand{\doi}[1]{doi:\urlalt{http://dx.doi.org/#1}{#1}}
\providecommand{\bibinfo}[2]{#2}

\bibitemdeclare{inproceedings}{Augustsson1985}
\bibitem{Augustsson1985}
\bibinfo{author}{Lennart \surnamestart Augustsson\surnameend}
  (\bibinfo{year}{1985}): \emph{\bibinfo{title}{Compiling pattern matching}}.
\newblock In: {\sl \bibinfo{booktitle}{Functional Programming Languages and
  Computer Architecture}}, {\sl \bibinfo{series}{Lecture Notes in Computer
  Science}} \bibinfo{volume}{201}, \bibinfo{publisher}{Springer-Verlag}, pp.
  \bibinfo{pages}{368--381}, \doi{10.1007/3-540-15975-4_48}.

\bibitemdeclare{book}{Box1970}
\bibitem{Box1970}
\bibinfo{author}{George~E.P. \surnamestart Box\surnameend} \&
  \bibinfo{author}{Gwilym~M. \surnamestart Jenkins\surnameend}
  (\bibinfo{year}{1970}): \emph{\bibinfo{title}{Time series analysis:
  Forecasting and control}}.
\newblock \bibinfo{publisher}{Holden--Day}, \bibinfo{address}{San Francisco}.

\bibitemdeclare{article}{Broy1988}
\bibitem{Broy1988}
\bibinfo{author}{Manfred \surnamestart Broy\surnameend} (\bibinfo{year}{1988}):
  \emph{\bibinfo{title}{Nondeterministic data flow programs: how to avoid the
  merge anomaly}}.
\newblock {\sl \bibinfo{journal}{Science of Computer Programming}}
  \bibinfo{volume}{10}, pp. \bibinfo{pages}{65--85},
  \doi{10.1016/0167-6423(88)90016-0}.

\bibitemdeclare{article}{Broy2001}
\bibitem{Broy2001}
\bibinfo{author}{Manfred \surnamestart Broy\surnameend} \&
  \bibinfo{author}{Gheorghe \surnamestart Ştefănescu\surnameend}
  (\bibinfo{year}{2001}): \emph{\bibinfo{title}{The algebra of stream
  processing functions}}.
\newblock {\sl \bibinfo{journal}{Theoretical Computer Science}}
  \bibinfo{volume}{258}(\bibinfo{number}{1--2}), pp. \bibinfo{pages}{99--129},
  \doi{10.1016/S0304-3975(99)00322-9}.

\bibitemdeclare{inproceedings}{Caspi1987}
\bibitem{Caspi1987}
\bibinfo{author}{Paul \surnamestart Caspi\surnameend}, \bibinfo{author}{Daniel
  \surnamestart Pilaud\surnameend}, \bibinfo{author}{Nicolas \surnamestart
  Halbwachs\surnameend} \& \bibinfo{author}{John \surnamestart
  Plaice\surnameend} (\bibinfo{year}{1987}): \emph{\bibinfo{title}{Lustre: A
  Declarative Language for Programming Synchronous Systems}}.
\newblock In: {\sl \bibinfo{booktitle}{POPL}}, \bibinfo{publisher}{ACM Press},
  pp. \bibinfo{pages}{178--188}, \doi{10.1145/41625.41641}.

\bibitemdeclare{article}{Cytron1991}
\bibitem{Cytron1991}
\bibinfo{author}{Ron \surnamestart Cytron\surnameend}, \bibinfo{author}{Jeanne
  \surnamestart Ferrante\surnameend}, \bibinfo{author}{Barry~K. \surnamestart
  Rosen\surnameend}, \bibinfo{author}{Mark~N. \surnamestart Wegman\surnameend},
  \bibinfo{author}{\surnamestart \surnameend} \& \bibinfo{author}{F.~Kenneth
  \surnamestart Zadeck\surnameend} (\bibinfo{year}{1991}):
  \emph{\bibinfo{title}{Efficiently computing static single assignment form and
  the control dependence graph}}.
\newblock {\sl \bibinfo{journal}{TOPLAS}}
  \bibinfo{volume}{13}(\bibinfo{number}{4}), pp. \bibinfo{pages}{451--490},
  \doi{10.1145/115372.115320}.

\bibitemdeclare{manual}{scade}
\bibitem{scade}
\bibinfo{organization}{Esterel Technologies} (\bibinfo{year}{2013}):
  \emph{\bibinfo{title}{Scade Suite}}.
\newblock
  \urlprefix\url{http://www.esterel-technologies.com/products/scade-suite/}.

\bibitemdeclare{inproceedings}{Henzinger1996}
\bibitem{Henzinger1996}
\bibinfo{author}{Thomas~A. \surnamestart Henzinger\surnameend}
  (\bibinfo{year}{1996}): \emph{\bibinfo{title}{The Theory of Hybrid
  Automata}}.
\newblock In: {\sl \bibinfo{booktitle}{Proc. Logic in Computer Science (LICS
  '96)}}, \bibinfo{publisher}{IEEE Computer Society}, pp.
  \bibinfo{pages}{278--292}, \doi{10.1109/LICS.1996.561342}.

\bibitemdeclare{article}{Hughes2000}
\bibitem{Hughes2000}
\bibinfo{author}{John \surnamestart Hughes\surnameend} (\bibinfo{year}{2000}):
  \emph{\bibinfo{title}{Generalising monads to arrows}}.
\newblock {\sl \bibinfo{journal}{Sci. Comput. Program.}}
  \bibinfo{volume}{37}(\bibinfo{number}{1-3}), pp. \bibinfo{pages}{67--111},
  \doi{10.1016/S0167-6423(99)00023-4}.

\bibitemdeclare{manual}{simulink}
\bibitem{simulink}
\bibinfo{organization}{The MathWorks} (\bibinfo{year}{2000}):
  \emph{\bibinfo{title}{Simulink, Dynamic System Simulation for Matlab ---
  Using Simulink}}.
\newblock \urlprefix\url{http://www.mathworks.com}.

\bibitemdeclare{article}{Moggi1991}
\bibitem{Moggi1991}
\bibinfo{author}{Eugenio \surnamestart Moggi\surnameend}
  (\bibinfo{year}{1991}): \emph{\bibinfo{title}{Notions of Computation and
  Monads}}.
\newblock {\sl \bibinfo{journal}{Information and Computation}}
  \bibinfo{volume}{93}(\bibinfo{number}{1}), pp. \bibinfo{pages}{55--92},
  \doi{10.1016/0890-5401(91)90052-4}.

\bibitemdeclare{inproceedings}{Nilsson2002}
\bibitem{Nilsson2002}
\bibinfo{author}{Henrik \surnamestart Nilsson\surnameend},
  \bibinfo{author}{Antony \surnamestart Courtney\surnameend} \&
  \bibinfo{author}{John \surnamestart Peterson\surnameend}
  (\bibinfo{year}{2002}): \emph{\bibinfo{title}{Functional Reactive
  Programming, Continued}}.
\newblock In: {\sl \bibinfo{booktitle}{Haskell Workshop}},
  \bibinfo{publisher}{ACM}, pp. \bibinfo{pages}{51--64},
  \doi{10.1145/581690.581695}.

\bibitemdeclare{article}{Niqui2013}
\bibitem{Niqui2013}
\bibinfo{author}{Milad \surnamestart Niqui\surnameend} \& \bibinfo{author}{Jan
  \surnamestart Rutten\surnameend} (\bibinfo{year}{2013}):
  \emph{\bibinfo{title}{Stream processing coalgebraically}}.
\newblock {\sl \bibinfo{journal}{Science of Computer Programming}}
  \bibinfo{volume}{78}, pp. \bibinfo{pages}{2192--2215},
  \doi{10.1016/j.scico.2012.07.013}.

\bibitemdeclare{article}{Pardo1998}
\bibitem{Pardo1998}
\bibinfo{author}{Alberto \surnamestart Pardo\surnameend}
  (\bibinfo{year}{1998}): \emph{\bibinfo{title}{Monadic Corecursion --
  Definition, Fusion Laws and Applications}}.
\newblock {\sl \bibinfo{journal}{ENTCS}} \bibinfo{volume}{11}, pp.
  \bibinfo{pages}{105--139}, \doi{10.1016/S1571-0661(04)00055-6}.

\bibitemdeclare{misc}{simulink-params}
\bibitem{simulink-params}
\bibinfo{author}{Guy \surnamestart Rouleau\surnameend} \& \bibinfo{author}{Seth
  \surnamestart Popinchalk\surnameend} (\bibinfo{year}{2008}):
  \emph{\bibinfo{title}{Initializing Parameters}}.
\newblock \bibinfo{howpublished}{Matlab Central Blog}.
\newblock
  \urlprefix\url{http://blogs.mathworks.com/seth/2008/12/25/initializing-parameters/}.
\newblock \bibinfo{note}{Retrieved 2013-12-31}.

\bibitemdeclare{article}{Rutten2000}
\bibitem{Rutten2000}
\bibinfo{author}{Jan~J.M.M. \surnamestart Rutten\surnameend}
  (\bibinfo{year}{2000}): \emph{\bibinfo{title}{Universal coalgebra: a theory
  of systems}}.
\newblock {\sl \bibinfo{journal}{Theoretical Computer Science}}
  \bibinfo{volume}{249}(\bibinfo{number}{1}), pp. \bibinfo{pages}{3--80},
  \doi{10.1016/S0304-3975(00)00056-6}.

\bibitemdeclare{book}{Spivey1988}
\bibitem{Spivey1988}
\bibinfo{author}{J.~M. \surnamestart Spivey\surnameend} (\bibinfo{year}{1988}):
  \emph{\bibinfo{title}{The Z Notation: a reference manual}}.
\newblock \bibinfo{series}{International Series in Computer Science},
  \bibinfo{publisher}{Prentice Hall}.

\bibitemdeclare{manual}{SupercolliderHome}
\bibitem{SupercolliderHome}
 (\bibinfo{year}{2011}): \emph{\bibinfo{title}{{S}upercollider {H}omepage}}.
\newblock \urlprefix\url{http://supercollider.sourceforge.net/}.

\bibitemdeclare{incollection}{Telford1997}
\bibitem{Telford1997}
\bibinfo{author}{Alastair \surnamestart Telford\surnameend} \&
  \bibinfo{author}{David~A. \surnamestart Turner\surnameend}
  (\bibinfo{year}{1997}): \emph{\bibinfo{title}{Ensuring streams flow}}.
\newblock In \bibinfo{editor}{Michael \surnamestart Johnson\surnameend},
  editor: {\sl \bibinfo{booktitle}{Algebraic Methodology and Software
  Technology}}, {\sl \bibinfo{series}{Lecture Notes in Computer Science}}
  \bibinfo{volume}{1349}, \bibinfo{publisher}{Springer-Verlag}, pp.
  \bibinfo{pages}{509--523}, \doi{10.1007/BFb0000493}.

\bibitemdeclare{article}{Turner2004}
\bibitem{Turner2004}
\bibinfo{author}{David~A. \surnamestart Turner\surnameend}
  (\bibinfo{year}{2004}): \emph{\bibinfo{title}{Total Functional Programming}}.
\newblock {\sl \bibinfo{journal}{Univ. Comput. Sci.}}
  \bibinfo{volume}{10}(\bibinfo{number}{7}), pp. \bibinfo{pages}{751--768},
  \doi{10.3217/jucs-010-07-0751}.

\bibitemdeclare{article}{Wan2000}
\bibitem{Wan2000}
\bibinfo{author}{Zhanyong \surnamestart Wan\surnameend} \&
  \bibinfo{author}{Paul \surnamestart Hudak\surnameend} (\bibinfo{year}{2000}):
  \emph{\bibinfo{title}{Functional reactive programming from first
  principles}}.
\newblock {\sl \bibinfo{journal}{SIGPLAN Not.}}
  \bibinfo{volume}{35}(\bibinfo{number}{5}), pp. \bibinfo{pages}{242--252},
  \doi{10.1145/358438.349331}.

\bibitemdeclare{book}{SupercolliderBook}
\bibitem{SupercolliderBook}
\bibinfo{author}{Scott \surnamestart Wilson\surnameend}, \bibinfo{author}{David
  \surnamestart Cottle\surnameend} \& \bibinfo{author}{Nick \surnamestart
  Collins\surnameend} (\bibinfo{year}{2011}): \emph{\bibinfo{title}{{T}he
  {S}upercollider {B}ook}}.
\newblock \bibinfo{publisher}{The MIT Press}.
\newblock \urlprefix\url{http://supercolliderbook.net}.

\end{thebibliography}

\end{document}